\documentclass[a4paper,11pt]{article}
\usepackage{jcappub} 
\usepackage{soul}
\usepackage{subcaption}
\usepackage{ulem}
\usepackage{booktabs} 
\usepackage{makecell}

\title{Anisotropic cosmic birefringence estimates for forthcoming CMB experiments}

\author[a]{M. Longo Minnolo, }

\author[a]{M. Billi,}
\emailAdd{longo9855@gmail.com, matteo.billi@dfa.unict.it, giuseppe.puglisi2@unict.it}
\author[a,b,c]{G.  Puglisi}
\affiliation[a]{Dipartimento di Fisica e Astronomia, Universit\`a degli Studi di Catania, via S. Sofia, 64, 95123, Catania, Italy}
\affiliation[b]{INFN - Sezione di Catania, Via S. Sofia 64, 95123 Catania, Italy}
\affiliation[c]{INAF - Osservatorio Astrofisico di Catania, via S. Sofia 78, 95123 Catania, Italy}
\abstract{Cosmic Birefringence (CB) is a unique probe of physics beyond the Standard Model, potentially sourced by an axion-like field coupled to photons via a Chern-Simons interaction. Recent Planck analyses have strengthened evidence for a non-zero isotropic rotation angle, yet its anisotropic component remains undetected, despite being the key to distinguishing a homogeneous quintessence-like field from spatially varying axion-like dark matter. We forecast the sensitivity of forthcoming ground- and space-based CMB experiments to anisotropic CB, applying a patch-based localization technique to simulated CMB-plus-noise maps with white noise levels representative of Simons Observatory SAT- and LiteBIRD-like 
sensitivities. Local rotation angles are reconstructed on independent 
sky patches by minimizing the $\chi^2_{EB}$ statistic built from the $EB$ D-estimator. 
Under our idealized simulations, the forecast angular power spectrum improves current Planck-based constraints on the CB anisotropy statistical uncertainty by a factor of 26-40 for a satellite-like mission and 11-15 for a realistic ground-based footprint. Translating these into constraints on the scale-invariant amplitude, fitting the reconstructed bandpowers with a scale-invariant amplitude model, we forecast with a 95 \% C.L. $A_{\rm CB} < 3.5\times10^{-4}\,{\rm deg}^2$ for the space-like mission and $A_{\rm CB} < 7.1\times10^{-4}\,{\rm deg}^2$ for the ground-based footprint, improving on the current full-sky Planck constraint by a factor of about 20 and 10, respectively. These values represent upper bounds on the achievable gain, and indicate that forthcoming CMB experiments will be able to further shed light on anisotropic CB.}

\begin{document}
\maketitle
\flushbottom

\section{Introduction}
\label{sec:intro}

Symmetries play a foundational role in the construction of the Standard Model of particle physics (SM). In this model, electromagnetic interaction preserves parity, but in weak interaction, parity violation is a well-established and experimentally verified phenomenon, originally proposed in~\cite{Lee1956} and later confirmed in~\cite{Wu1957}. However, the $\Lambda$CDM model, demands the existence of new physics beyond the SM. 

Specifically, the SM fails to provide viable candidates for the nature of cold dark matter and dark energy, which together constitute the vast majority of the Universe's energy budget~\cite{Planck2018_CosmologicalParameters}. Many beyond the SM 
theories introduce a new pseudoscalar field $\chi$, which changes sign under the inversion of spatial coordinates. Such fields frequently emerge in high-energy physics, string theory compactifications, and theories addressing the strong-CP problem of Quantum Chromodynamics (see e.g.~\cite{Peccei1977,Weinberg1978,Marsh2016_AxionCosmology}). 

Generically referred to as axions or axion-like particles, they are often associated with the spontaneous breaking of a global $U(1)$ symmetry at a high energy scale $f_a$, known as the axion decay constant~\cite{RevModPhys.82.557}. From a cosmological perspective, a rolling or oscillating pseudoscalar field,$\chi$ represents a good candidate to act as quintessence, driving the late-time accelerated expansion of the Universe, or as the cold dark matter itself~\cite{Marsh2016_AxionCosmology}. If the fundamental physics governing the dark energy or the dark matter violates parity, $\chi$  can couple directly to the electromagnetic field and  the Lagrangian density is  modified by the addition of a parity-violating Chern-Simons (CS) interaction term~\cite{Carroll1990_Lorentz} as follows:

\begin{equation}
    \mathcal{L}_{EM} +  \mathcal{L}_{CS}  = -\frac{1}{4}F_{\mu\nu}F^{\mu\nu} + \frac{1}{4} g_{\chi\gamma} \chi F_{\mu\nu} \tilde{F}^{\mu\nu}, 
    \label{eq:sm_em_lag}
\end{equation}
where $F_{\mu\nu}$ is the standard electromagnetic field tensor, $g_{\chi\gamma}$ is the coupling constant and $\tilde{F}^{\mu\nu} = \frac{1}{2}\epsilon^{\mu\nu\rho\sigma}F_{\rho\sigma}$ is the dual tensor of the the electromagnetic field tensor, with $\epsilon^{\mu\nu\rho\sigma}$ being the four-dimensional Levi-Civita symbol~\cite{Marsh2016_AxionCosmology}. 
Two distinct realizations of this coupling are discussed in the
literature~\cite{Carroll1990_Lorentz}: either $\chi$ is replaced by a fixed,
non-dynamical four-vector, which explicitly introduces a preferred direction
in space-time and spontaneously breaks Lorentz and CPT symmetry; or $\chi$ is
a genuinely dynamical pseudoscalar field, whose background time evolution
sources the effect discussed below. In this work we adopt the latter
interpretation, treating $\chi$ as a dynamical field.

As CMB photons travel for over 13 billion years from the surface of last
scattering through this varying background field, the coupling induces a
cumulative modification in their propagation~\cite{Lue1999}. In particular,
the phase velocities of left- and right-handed circularly polarized photons
accumulate. This phenomenon results in the in-vacuo rotation of the plane of
linear polarization of the CMB photons, an effect fundamentally known as
{Cosmic Birefringence (CB)}. As $\chi$ could be
inhomogeneously distributed, it is reasonable to model CB with a
direction-dependent set of angles $\alpha(\hat{n})$\footnote{Only when the
field is exclusively time dependent, the effect reduce to its isotropic
form, which is parameterized by a single, global rotation $\alpha$.}.

While there are hints of an isotropic cosmic birefringence detection at approximately  the $3\sigma$ level from self-calibrated \textit{WMAP} and \textit{Planck} data, with $\bar{\alpha} = 0.342\text{°} \, {}^{+0.094\text{°}}_{-0.091\text{°}}$~\cite{Minami:2020fin,PhysRevLett.128.091302,PhysRevD.106.063503}, that more recently have been strengthened by the independent 
confirmation with \textit{ACT DR6}~\cite{ACT,pbc3-t52s}, reaching $\bar{\alpha} = 0.277\text{°} \pm 0.057\text{°}$ at $4.8\sigma$ when combining datasets~\cite{hcsm-39qw}, no significant signal of anisotropic cosmic birefringence has been detected to date. Current observations remain fully compatible with a null effect, with the most stringent upper limits on the scale-invariant amplitude $A_{CB} = L(L+1)C_L^{\alpha\alpha}/(2\pi)$ set to $A_{CB} < 0.007 \, \text{deg}^2$ from full-sky \textit{Planck} data~\cite{Bortolami2022}, $A_{CB}< 0.044\times10^{-4} \, \text{deg}^2$ from \textit{BICEP/Keck}~\cite{Bicepkeck}, $A_{CB} < 0.15\times10^{-4} \, \text{deg}^2 \ $ from \textit{ACT DR6} ~\cite{kramer2026atacamacosmologytelescopeconstraints}, $A_{CB} < 0.53\times10^{-4} \, \text{deg}^2$ from \textit{SPT}~\cite{SPTbir} and $A_{CB} < 1.00 \times10^{-4} \, \text{deg}^2$ from combined datasets~\cite{Lonappan}.

The main goal of this work is to forecast the statistical sensitivity achievable by future CMB experiments (both ground- and space-based) to
anisotropic CB, by quantifying the expected error on the
anisotropic CB power spectrum under idealized simulation conditions. To achieve this, we employed the localization method introduced by~\cite{Gruppuso2020} and also utilized in~\cite{Bortolami2022}, whose results serve as a benchmark for this work. From the high-resolution simulated CMB maps, the methodological flow aims to build maps of the birefringence angle at low resolution, which allows us to evaluate the CB angular power spectrum on angular scales larger than approximately $7^\circ$.
The method consists in partition the full-sky map into sub-patches, obtained by adopting an HEALPix\footnote{https://healpix.sourceforge.net}~\cite{Gorski2005} pixelization scheme with a resolution of $N_{side}=8$. Within each individual patch, the birefringence angle is assumed to be isotropic and constant.

This paper is organized as follows: in section~\ref{sec:methodology} we describe the methodology used to extract the local rotation angle; in section~\ref{sec:dataandsim} we detail the features of the datasets and the simulated CMB maps employed for this study; in section~\ref{sec:pipeline} we outline the analysis pipeline applied to the selected sky patches; in section~\ref{sec:results} we present the results of this work; in section~\ref{sec:conclusion} we draw the conclusions.


\section{Methodology}
\label{sec:methodology}
The method consists in partition the full-sky map into sub-patches, obtained by adopting an HEALPix pixelization scheme with a resolution of $N_{side}=8$. Within each individual patch, the birefringence angle is assumed to be isotropic and constant.

Subsequently the D-estimators~\cite{PhysRevLett.102.161302}, are systematically applied to each of these sky patches. The detection of CB relies on measuring $TB$ and $EB$ cross-correlations, which are identically zero under the parity-conserving framework of standard electromagnetism~\cite{Lue1999,Finelli2009,Komatsu2022}. Although both correlations are valid theoretical indicators of parity
violation, $TB$ is intrinsically less constraining than $EB$, with  the $EB$ spectrum yielding a higher signal-to-noise ratio compared to the $TB$ counterpart. This is the reason why, in this work we focus only on the $EB$ correlation~\cite{Bortolami2022}.

Following~\cite{PhysRevLett.102.161302}, the $EB$ D-estimator is defined as:
\begin{equation}
    \hat{D}_\ell^{EB,\text{obs}}(\beta) = \hat{C}_\ell^{EB,\text{obs}} \cos(4\beta) - \frac{1}{2} \left( \hat{C}_\ell^{EE,\text{obs}} - \hat{C}_\ell^{BB,\text{obs}} \right) \sin(4\beta) \, ,
    \label{eq:D_estimator_data}
\end{equation}
whose expectation value vanishes precisely when $\beta$ equals the true rotation angle~\cite{Gruppuso_2012}. The local rotation angle $\beta$ in each patch is evaluated by minimizing the $\chi^2_{EB}$ statistic:

\begin{equation}
    \chi^2_{EB}(\beta) = \sum_{\ell,\ell'} \hat{D}^{EB,obs}_\ell(\beta) \left( M^{EB}_{\ell\ell'} \right)^{-1} \hat{D}^{EB,obs}_{\ell'}(\beta)
    \label{eq:chi_eb}
\end{equation}
where $\hat{D}^{EB,obs}_\ell(\beta)$ is the D-estimator obtained with the CMB spectra extracted in that specific patch, and $\left( M^{EB}_{\ell\ell'} \right)^{-1}$ is the inverse of the covariance matrix of the estimators. Since we employ this minimization following a frequentist approach based on simulated observations with a null intrinsic CB signal, the matrix $M^{EB}_{\ell\ell'}$ simply reduces to the EB covariance matrix computed for each patch. 

The systematic application of the D-estimators on these single portions of the sky allows for the reconstruction of map of the CB angle at a resolution of $N_{side}=8$. Finally, from this reconstructed map, the CB angular power spectrum  and the relative upper limits are derived.
\section{Dataset and simulation}
\label{sec:dataandsim}

\begin{figure}[htbp]
    \centering
    \begin{minipage}[t]{0.48\textwidth}
        \centering
        \textbf{Temperature mask} \\
        \includegraphics[width=\textwidth]{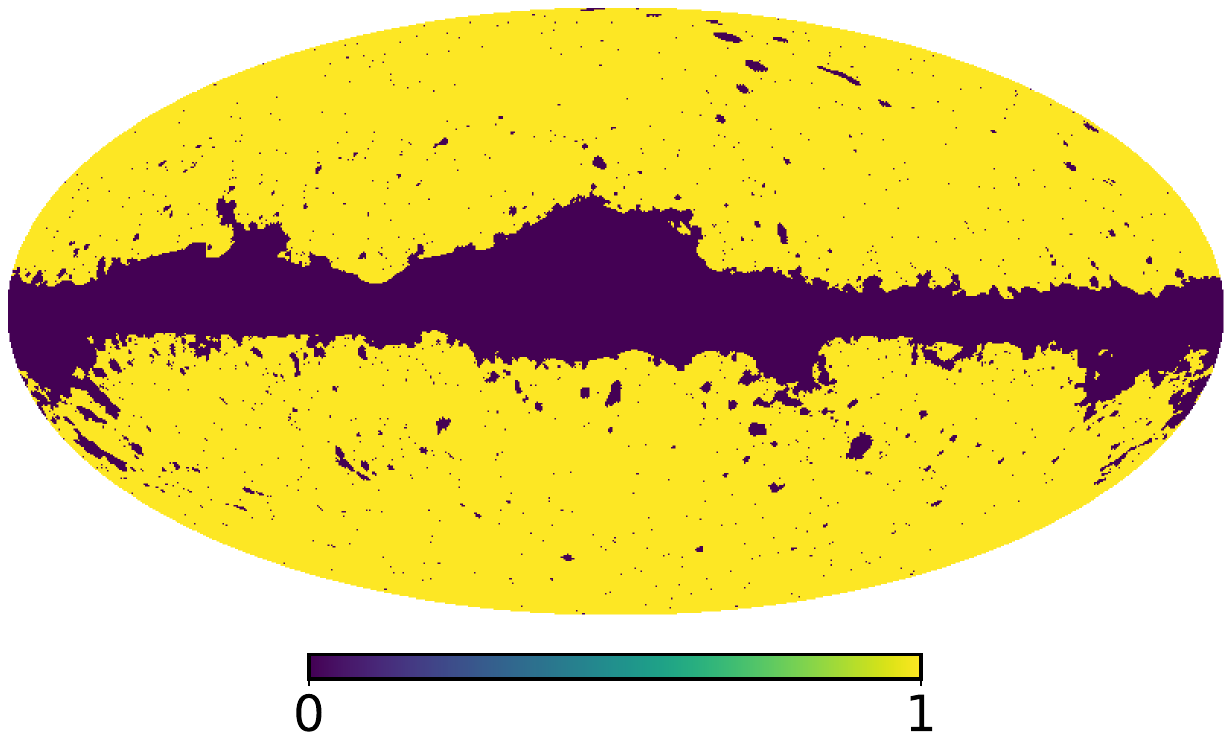}
    \end{minipage}
    \hfill
    \vspace{0.6cm}
    \begin{minipage}[t]{0.482\textwidth}
        \centering
        \textbf{Polarization mask} \\
        \includegraphics[width=\textwidth]{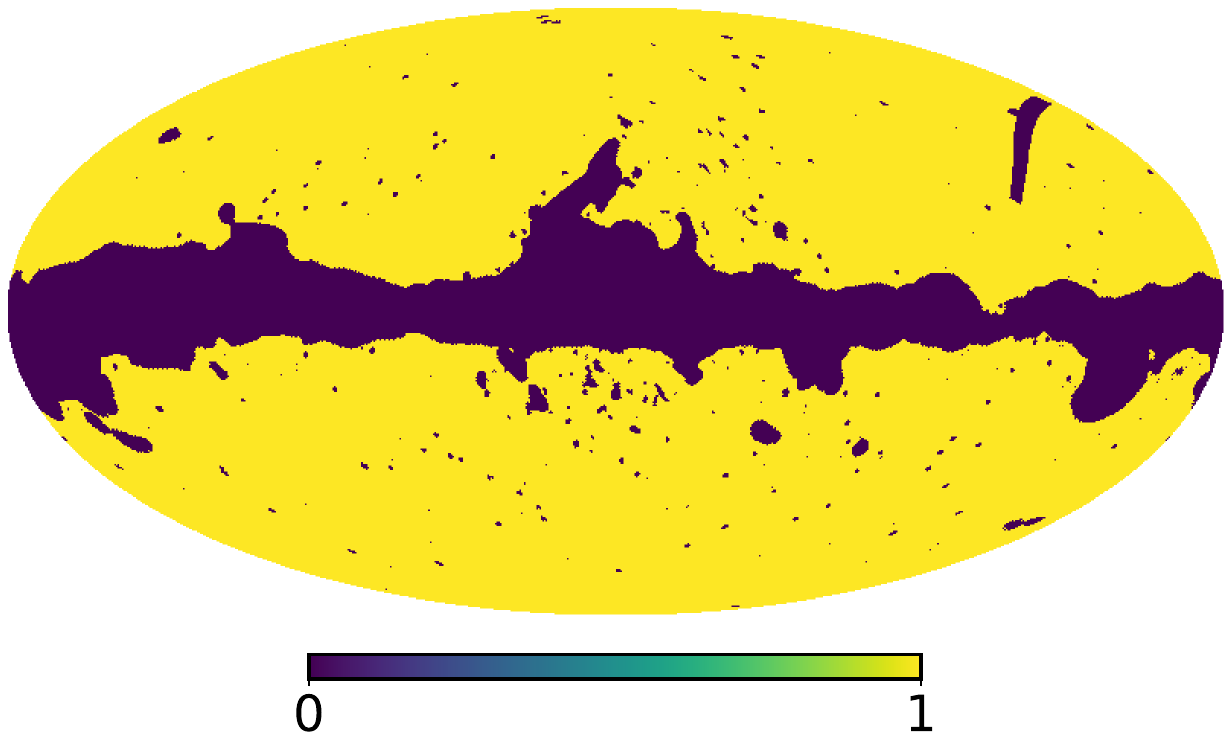}
    \end{minipage}
    \caption{Temperature (left) and polarization (right) PR3 masks used to reduce Galactic foreground contamination.}
    \label{fig:masks}
\end{figure}

For this analysis, we build two distinct sets of 1000 simulated \textit{CMB-plus-noise} maps at an \texttt{HEALPix} resolution $N_{\mathrm{side}}=512$, smoothed with a Gaussian beam with $16.5^\prime$ FWHM and the standard \texttt{HEALPix}  pixel window function. We employ \textit{Planck} fiducial cosmology~\cite{2020Planck3,2020Planck7} to  simulate 1000 CMB realizations in both temperature and polarization.
Then we combine the 1000 CMB simulated maps with two different sets of 1000 uncorrelated Gaussian noise maps, assuming a level of $2\,\mu\text{K-arcmin}$ for temperature and $\sqrt{2} \cdot 2\,\mu\text{K-arcmin}$ for polarization. These specific noise levels are representative of the expected sensitivities for current and forthcoming CMB experiments, such as the Simons Observatory Small Aperture Telescope (SO-SAT)~\cite{SimonsObservatory:2018koc} and LiteBIRD~\cite{LiteBIRD:2022cnt}. Each simulated map is then combined with \textit{Planck} 2018 Data Release (PR3) polarization and temperature masks (Figure \ref{fig:masks}), which leave approximately 78$\%$ of the sky uncovered, to exclude regions highly contaminated by galactic foregrounds.

To validate the pipeline used to generate the simulated maps, we
cross-correlate the two sets of CMB-plus-noise maps, which share the same
underlying 1000 CMB realizations but have statistically independent noise
realizations (see above). We estimate the $TT$, $EE$, $BB$,$TE$, $TB$, $EB$  angular power spectra within the multipole range $\ell \in [2, 2N_{\text{side}}]$. Because the two sets
share the same CMB sky, this cross-spectrum is unbiased by the noise power, enabling a direct comparison between the recovered spectra and the fiducial CMB model.

\begin{figure}[p]
	\centering
	\begin{subfigure}{0.48\textwidth}
		\centering
		\includegraphics[width=0.9\linewidth]{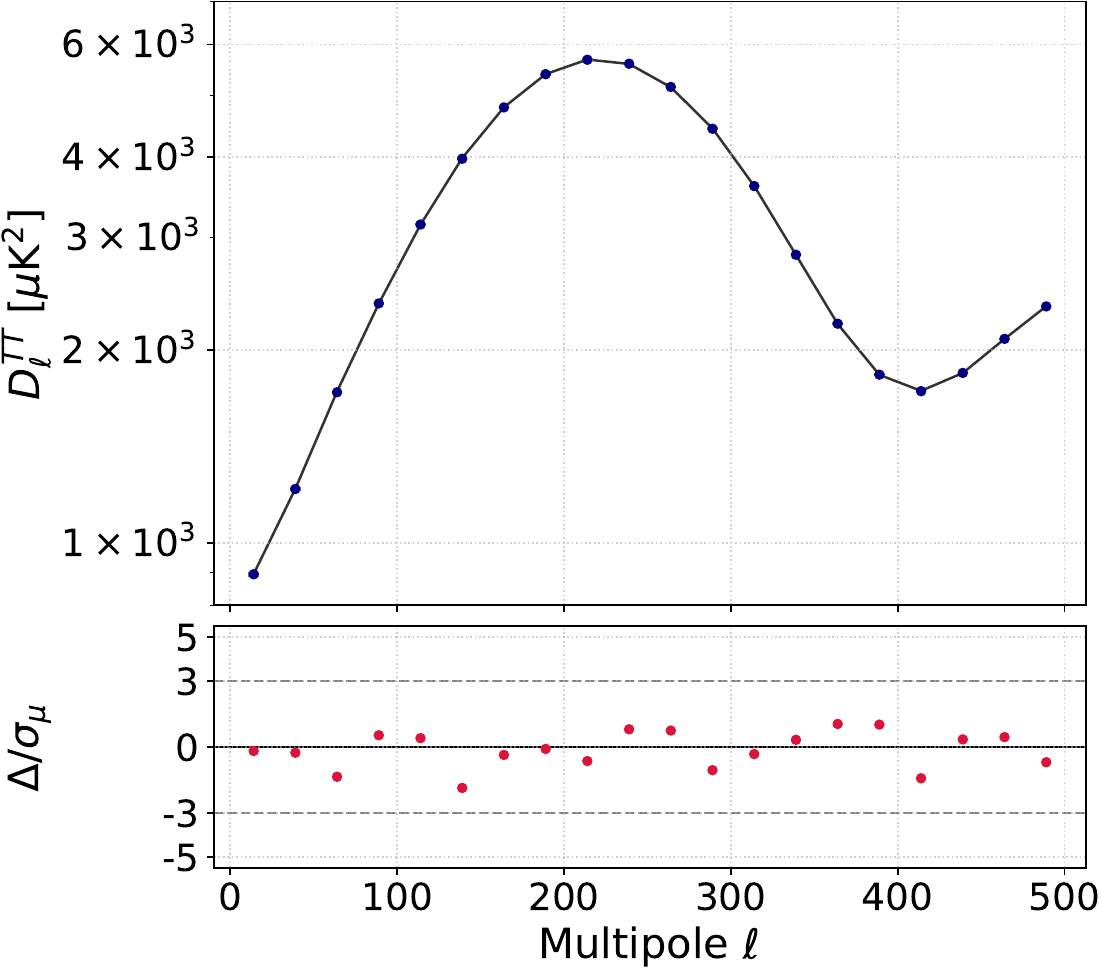}
		\label{fig:spec_tt}
	\end{subfigure}\hfill
	\begin{subfigure}{0.48\textwidth}
		\centering
		\includegraphics[width=0.9\linewidth]{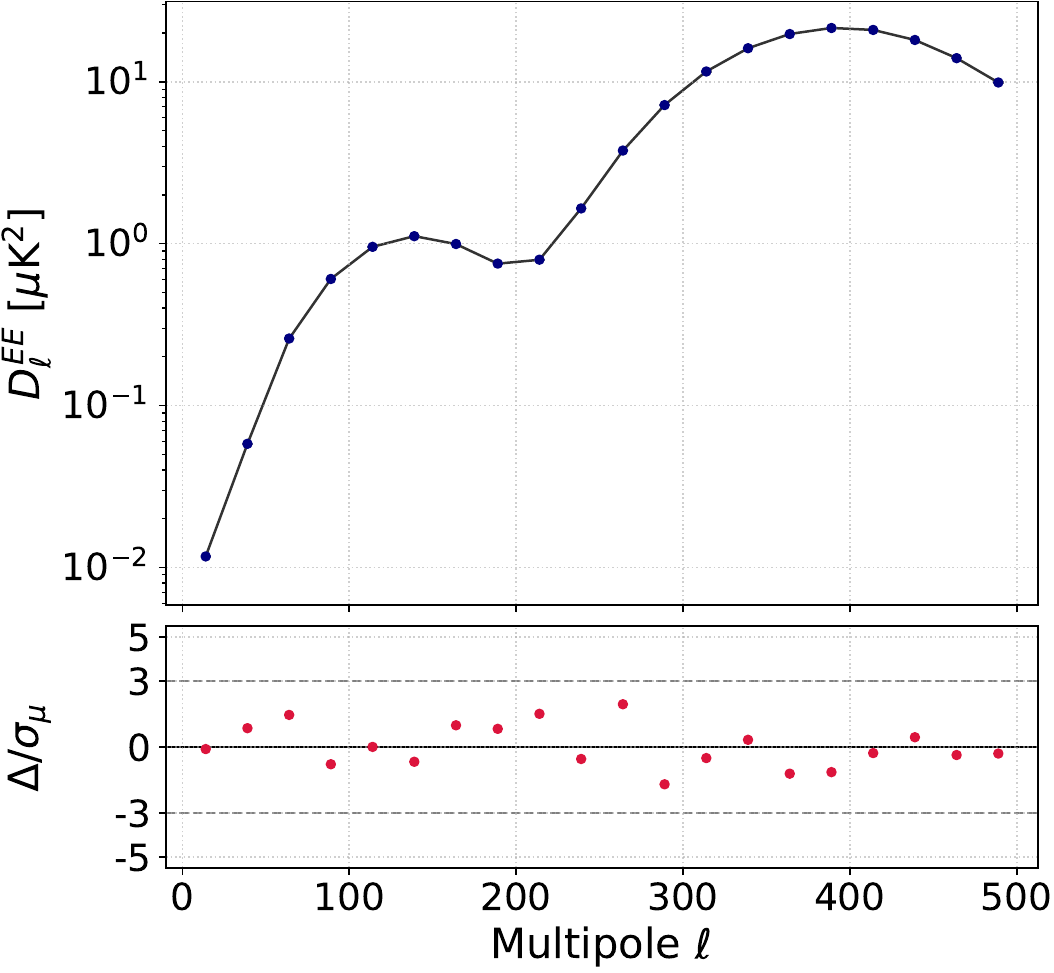}
		\label{fig:spec_ee}
	\end{subfigure}
	
	\vspace{0.4cm}
	
	\begin{subfigure}{0.48\textwidth}
		\centering
		\includegraphics[width=0.9\linewidth]{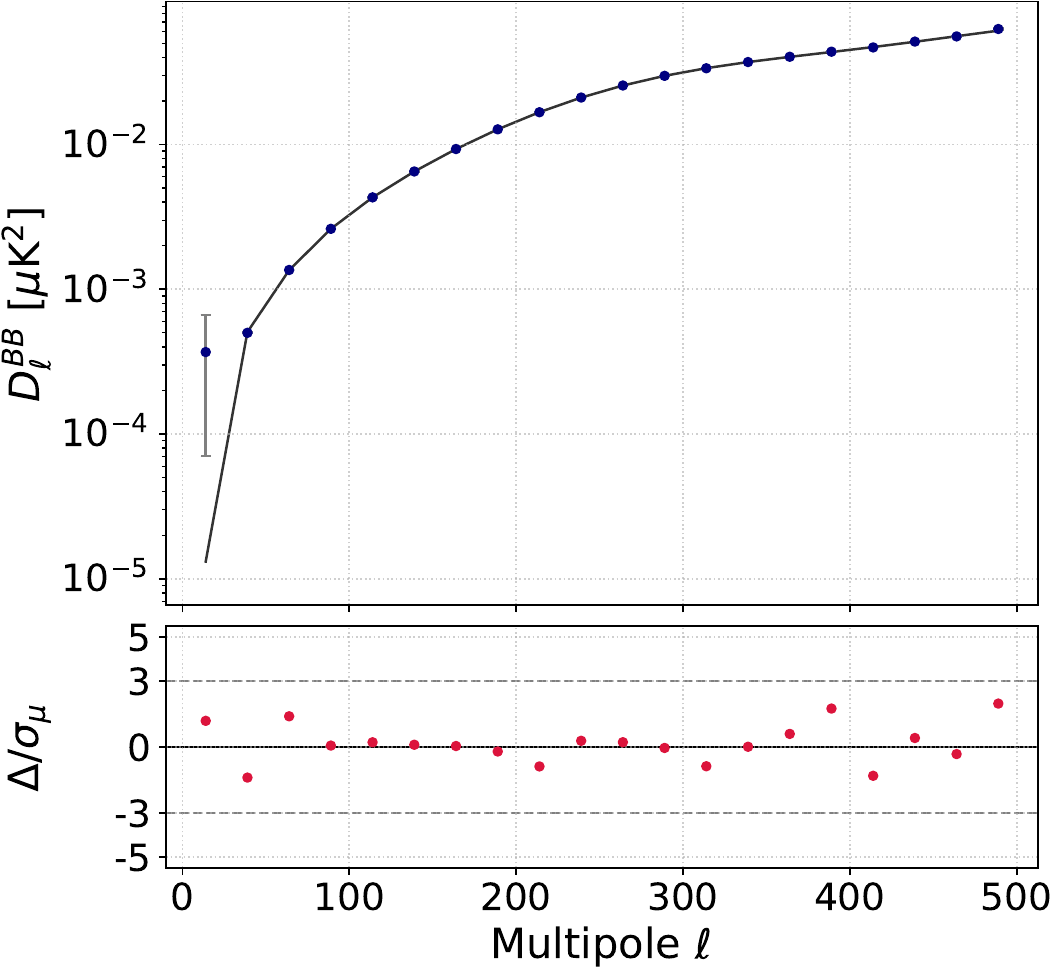}
		\label{fig:spec_bb}
	\end{subfigure}\hfill
	\begin{subfigure}{0.48\textwidth}
		\centering
		\includegraphics[width=0.9\linewidth]{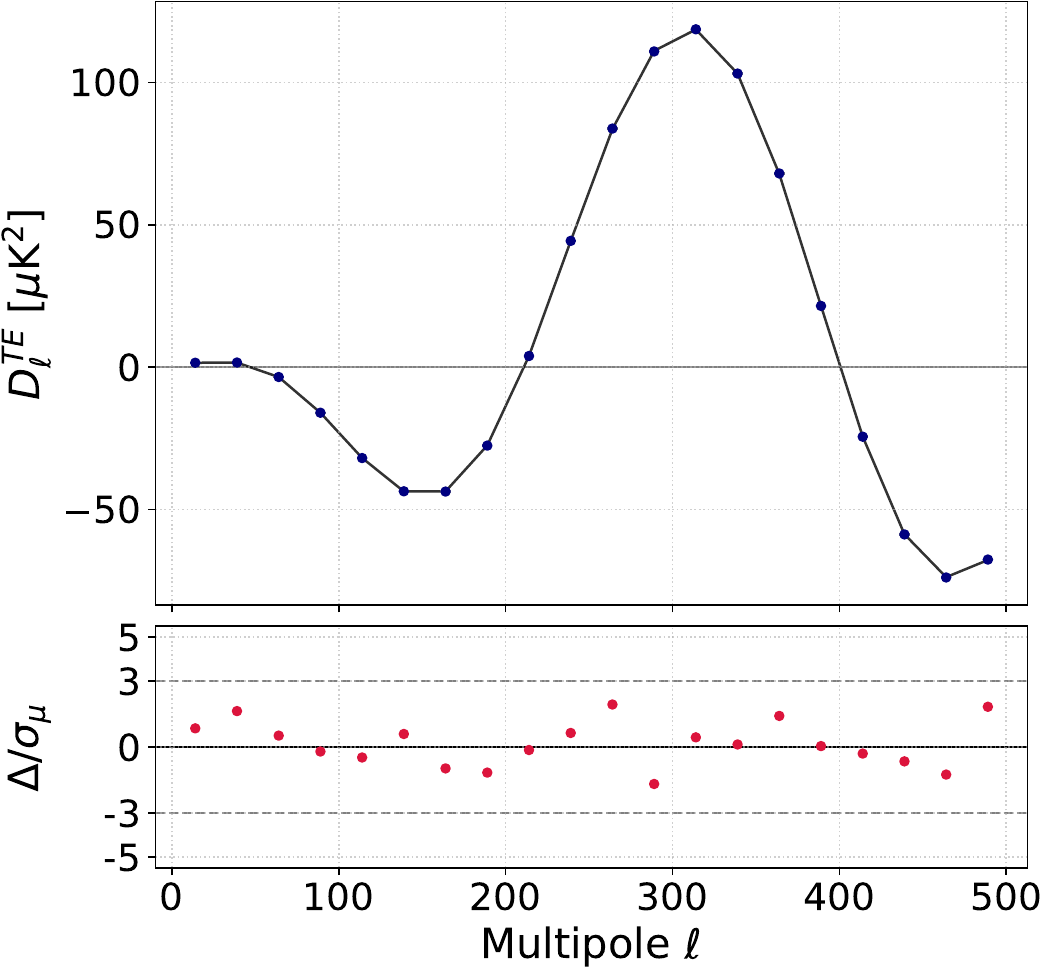}
		\label{fig:spec_te}
	\end{subfigure}
	
	\vspace{0.4cm}
	
	\begin{subfigure}{0.48\textwidth}
		\centering
		\includegraphics[width=0.9\linewidth]{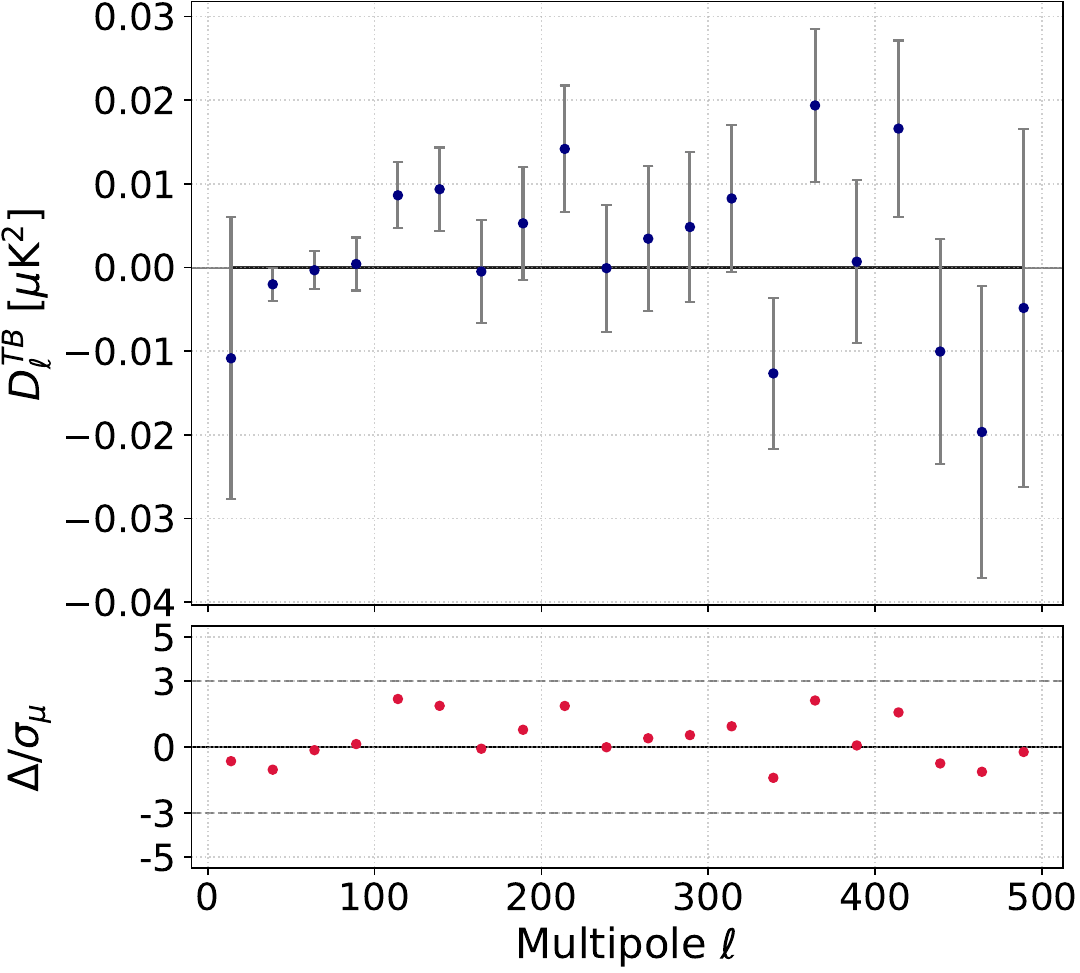}
		\label{fig:spec_tb}
	\end{subfigure}\hfill
	\begin{subfigure}{0.48\textwidth}
		\centering
		\includegraphics[width=0.9\linewidth]{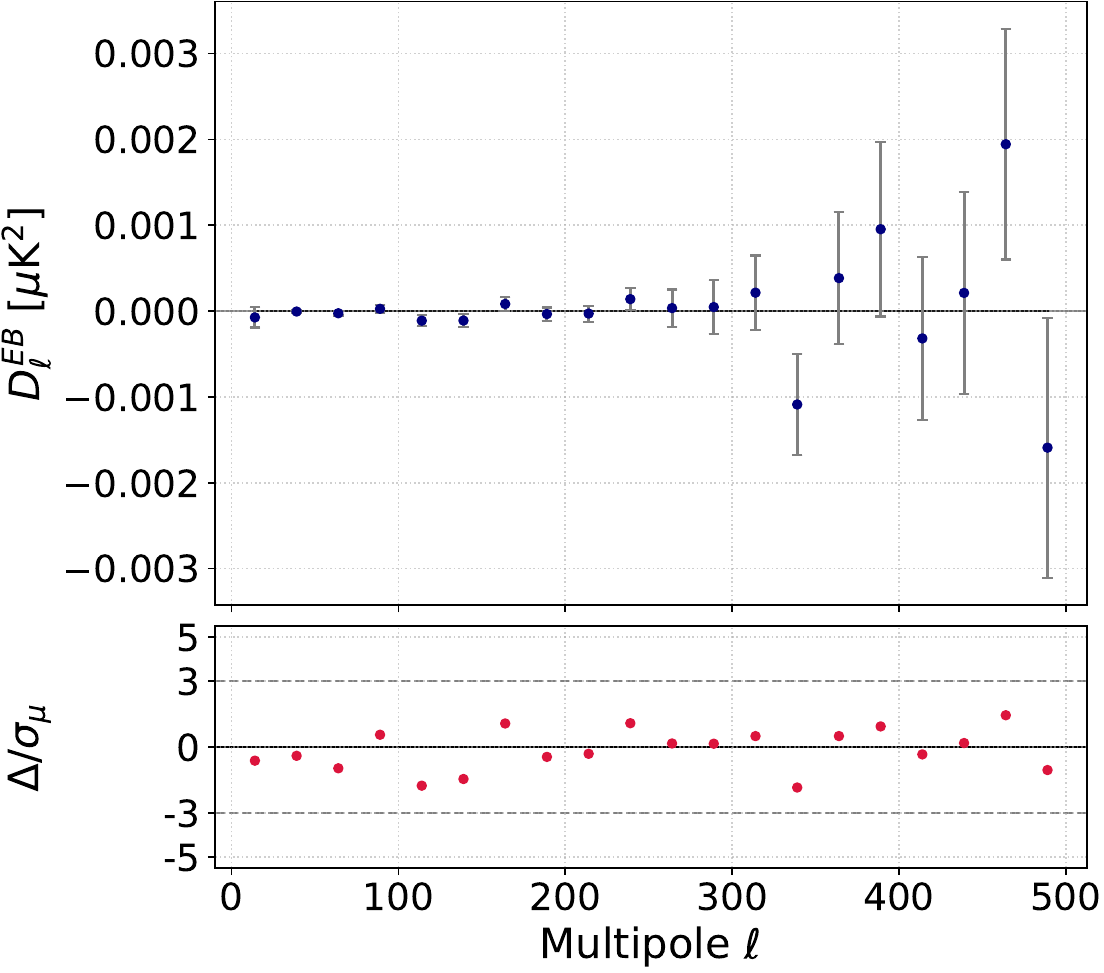}
		\label{fig:spec_eb}
	\end{subfigure}
	
	\caption{Pseudo-$C_\ell$ power spectra, in terms of the bandpowers $D_{\ell} = \ell(\ell+1)C_{\ell} /2\pi$, recovered from 1000 Monte Carlo simulations. The top panels show the mean recovered spectra (blue points) compared to the theoretical input (black line). The bottom panels display the deviation in units of the standard error, $\Delta / \sigma_\mu$, with the $\pm 3\sigma$ confidence intervals marked by dashed lines.}
	\label{fig:all_spectra_nside256}
\end{figure}

The spectra are computed using the Python package \texttt{pymaster}~\cite{Alonso2019}. Figure \ref{fig:all_spectra_nside256} shows the six spectra recovered from the 1000 Monte Carlo simulations, demonstrating that the simulation pipeline has been validated and does not introduce systematic bias into the measurement. 

\section{Analysis pipeline}
\label{sec:pipeline}
\begin{figure}[htbp]
	\centering
	\includegraphics[width=0.6\linewidth]{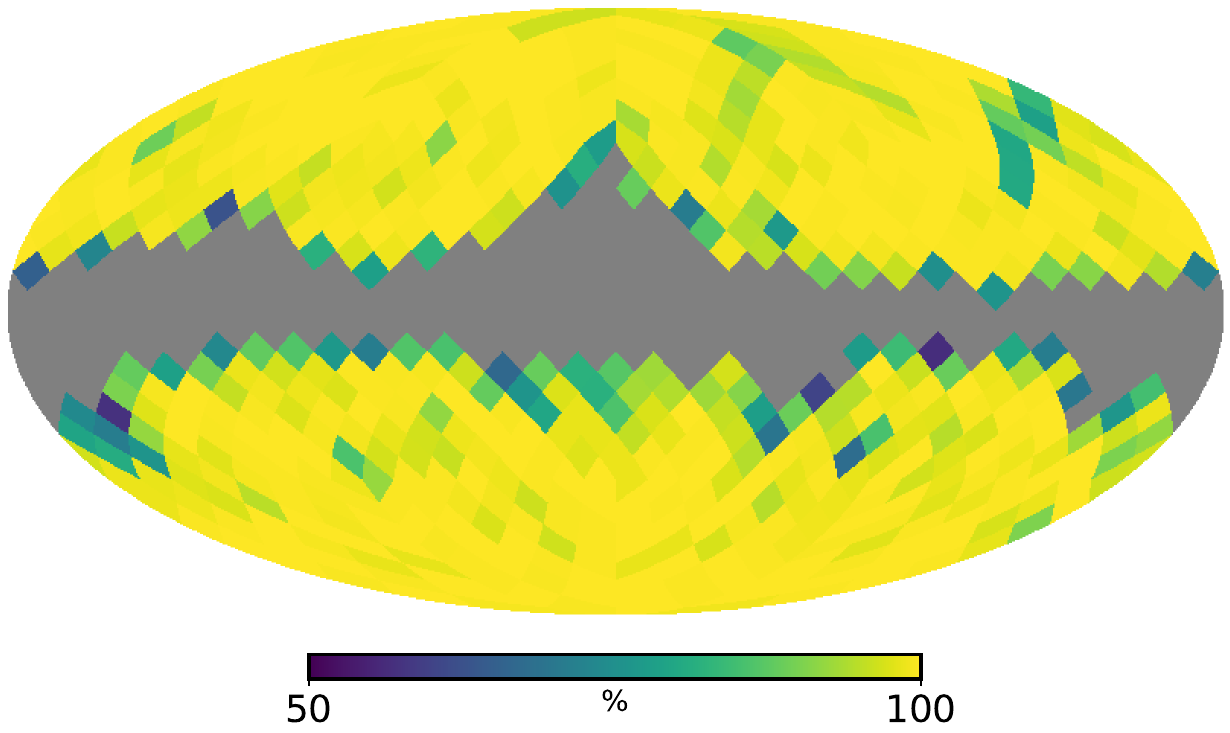}
	\caption{Percentage of active pixels for each patch. The gray patches have a fraction of active pixels  lower than 50\% and are therefore discarded from the analysis.}
	\label{fig:active_pixel}
\end{figure}

As outlined in previous sections, we apply the localization method using the sets of simulated map at resolution of $N_{\mathrm{side}}=512$ described above. The final mask applied to each map is the product of the Common \textit{Planck} temperature and polarization mask (see Figure~\ref{fig:masks}) and the specific mask defining the observed patch at $N_{\mathrm{side}}=8$. A large fraction of masked pixels within a single patch prevents the reliable extraction of CMB spectra. For this reason, we select and analyze only patches with at least 50\% of active pixels (Figure~\ref{fig:active_pixel}). 

This selection criterion yields a total of 592 patches. We apodize the final mask with a radius of $0.27^\circ$ and a $C^2$ profile \footnote{A $C^2$ apodization function smoothly transitions from 0 to 1 such that its first and second derivatives vanish at the boundaries. This level of smoothness avoids discontinuities in the second-order derivatives evaluated during the $B$-mode purification, minimizing numerical leakage.}
, in order to mitigate the $E-$to$-B$ leakage and cross-correlation between multipoles induced by the mask boundaries~\cite{Alonso2019}. 

We employ $B-$mode purification to properly correct for the $E-$to$-B$ leakage induced by the aggressive masking. The spectra are computed in the multipole range $\ell \in [50, 1024]$. The first 50 multipoles are excluded for two main reasons. First, at these large
angular scales, the small number of available modes makes the distribution
of the power spectrum estimator intrinsically non-Gaussian, regardless of whether a mask is applied. Second, and more relevant for our patch-based analysis, the mask breaks the statistical isotropy of the observed field at large scales, invalidating the diagonal approximation to the covariance matrix and significantly inflating the effective error bars~\cite{l50Couchot2017, l50Garcia2019, l50Planck}. 

Furthermore, the minimum $\ell$ multipole is physically motivated by the angular size of our $N_{\mathrm{side}}=8$ patches: although the theoretical minimum multipole for a $\approx 7.3^{\circ}$ region is $\ell \approx 25$, extracting power spectra near this fundamental limit introduces unacceptably large statistical errors. Extending the cut to $\ell \ge 50$ provides a necessary buffer, ensuring the analysis is restricted to modes with reliable measurements.
Finally, the power spectra are  binned using a constant width of $\Delta \ell=25$. To speed up computations, the pipeline has been parallelized using the \texttt{mpi4py} Python package. The simulations have been executed on 512 cores distributed across 4 nodes of the Perlmutter supercomputer  at  NERSC\footnote{https://www.nersc.gov/}. We estimate an overall computational cost of  approximately 240 node-hours.
\begin{figure}[htbp]
    \centering
    
    \begin{minipage}{0.48\textwidth}
        \centering
        \includegraphics[width=\textwidth]{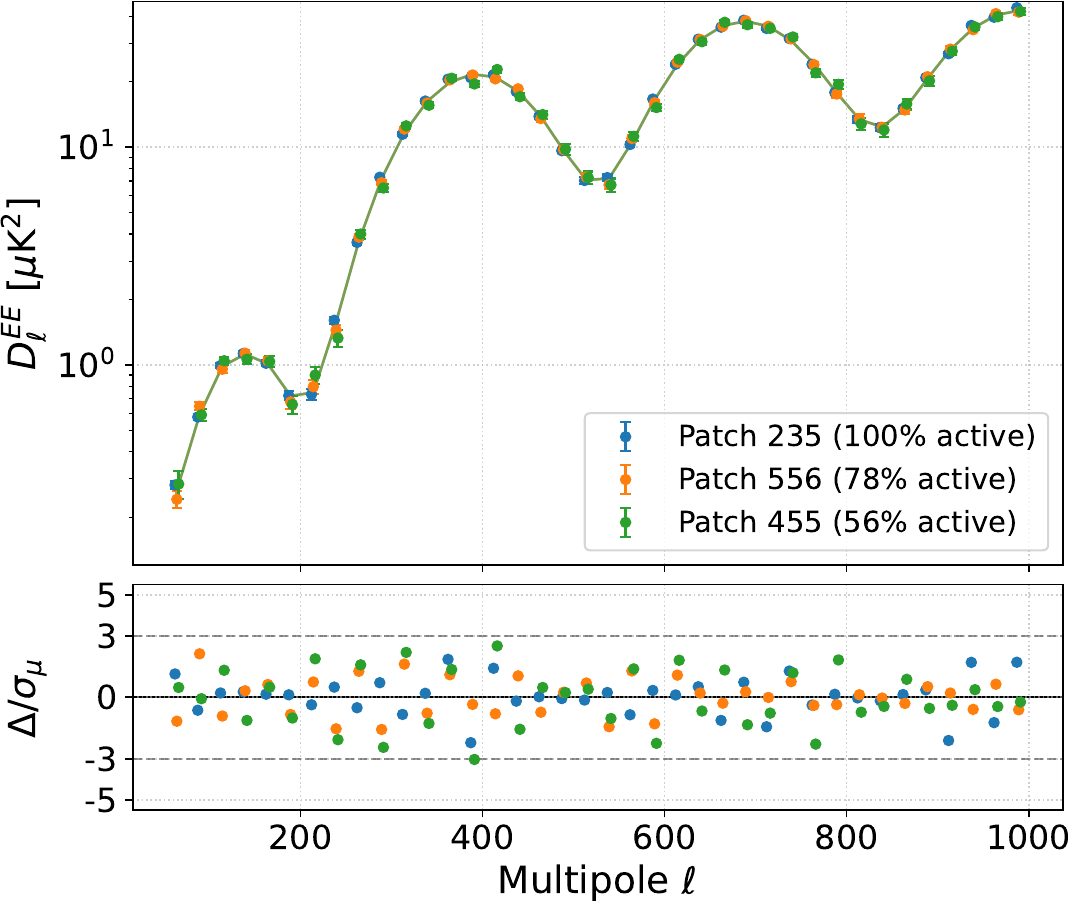}
    \end{minipage}\hfill
    \begin{minipage}{0.48\textwidth}
        \centering
        \includegraphics[width=\textwidth]{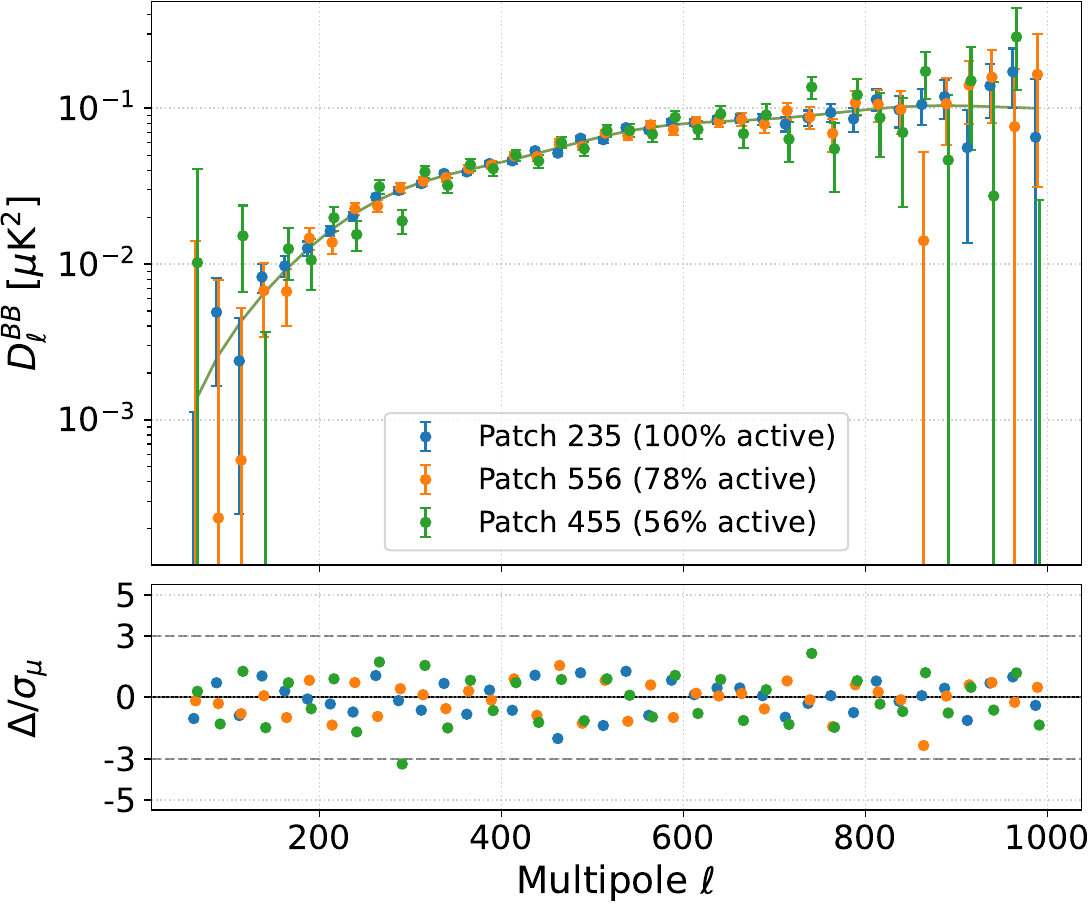}
    \end{minipage}
    
    \vspace{15pt} 
    
    \begin{minipage}{0.48\textwidth}
        \centering
        \includegraphics[width=\textwidth]{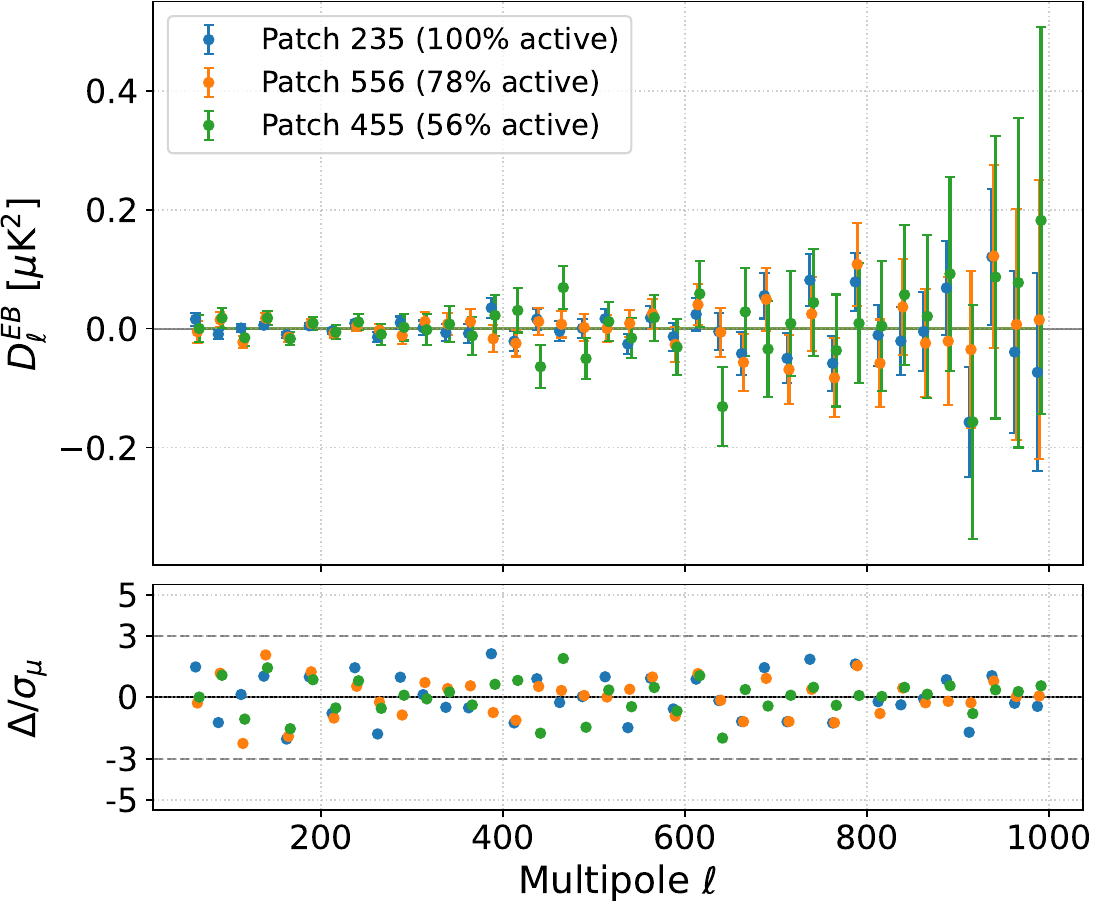}
    \end{minipage}
    
    \caption{Angular power spectra validation for three different sky fractions (active pixels: 100\%, 78\%, and 56\%). Top left: $EE$ power spectrum; Top right: $BB$ power spectrum; Bottom center: $EB$ power spectrum. Each plot displays the spectrum validation (top panel), and the deviation in units of the standard error, $\Delta / \sigma_\mu$ (bottom panel), with the $\pm 3\sigma$ confidence intervals marked by dashed lines.}
    \label{fig:combined_patch_analysis}
\end{figure}

Figure~\ref{fig:combined_patch_analysis}  shows the validation plots for the $EE$, $BB$, and $EB$ power spectra, expressed in terms of the bandpowers $D_{\ell} = \ell(\ell+1)C_{\ell} /2\pi$, for three representative patches. Notice that the selected patches exhibit different fractions of active pixels: 100\%, 77.9\%, and 56\%, respectively. However, it shows that the power spectra are correctly recovered and free from systematic biases.  

Covariance matrices for each patch are obtained using the analytical implementation provided by the \texttt{pymaster} library~\cite{Alonso2019}, which computes a Gaussian approximation of the pseudo-$C_\ell$ covariance from the mode-coupling matrix associated with each patch's apodized mask, combined with the fiducial theoretical CMB power spectra plus the assumed white noise level used to generate the simulations. This purely analytical approach follows the same methodology adopted in the reference framework of~\cite{Bortolami2022}, providing a computationally inexpensive and fully reproducible baseline covariance without the need for additional dedicated simulation sets.

\section{Results}
\label{sec:results}
\begin{figure}[htbp]
	\centering
	\vspace{2.0 cm}\includegraphics[width=0.7\linewidth]{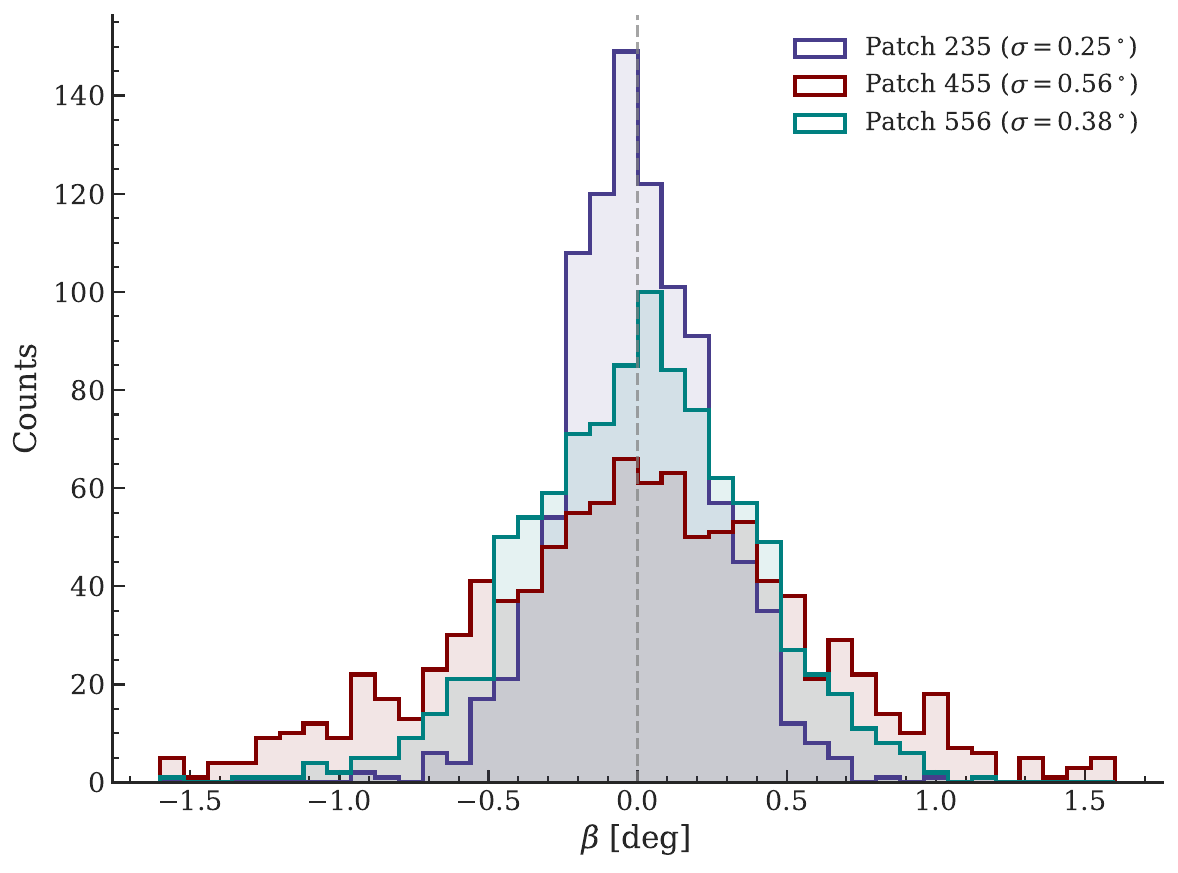}
	\caption{Distribution of the evaluated $\beta$ angles across 1000 simulations for three representative patches.}
	\label{fig:histcomparison}
\end{figure}

After obtaining the CMB power spectra and covariance matrices, the cosmic birefringence angle is estimated locally, assuming the effect to be isotropic within each patch. By minimizing the $\chi^2$(equation ~\ref{eq:chi_eb}) statistic, we estimate the CB angle $\beta$ within each patch across all Monte Carlo simulations. This procedure yield a set of 1000 birefringence angle maps. Figure~\ref{fig:histcomparison} shows the distribution of the $\beta$ angle across 1000 simulations for the three aforementioned patches. As expected, a lower fraction of active pixels widens the distribution, yet the estimations remain unbiased. 

\begin{figure}[htbp]
    \centering
    
    \includegraphics[width=0.6\linewidth]{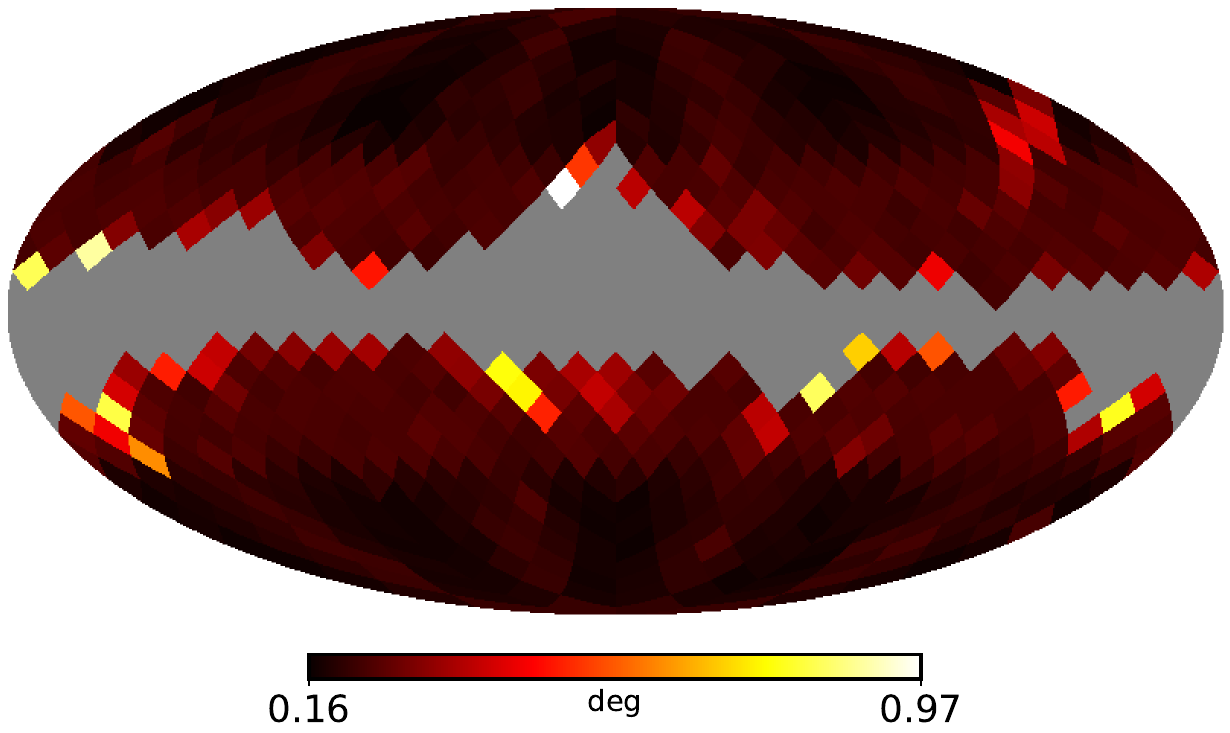}
    
    \vspace{0.4cm} 
    \caption{Standard deviation map $\sigma_{\beta}$ of the $\beta$ angles extracted from the 1000 MC simulations in this work, displayed with an adjusted dynamic colorbar to properly visually resolve the spatial variations. A direct comparison with the active pixel map (Figure~\ref{fig:active_pixel}), clearly indicates that higher values of $\sigma_{\beta}$ strongly correlate with heavily masked patches.}
    \label{fig:std_combined}
\end{figure}

Figure~\ref{fig:std_combined} shows the standard deviation map $\sigma_{\beta}$ of the $\beta$ angles extracted from the 1000 MC simulations in this work. Compared to the reference map presented by~\cite[Section 4, Figure 5, top right panel]{Bortolami2022}, the statistical uncertainties in this work are significantly lower, primarily due to the reduced noise levels in our simulated dataset. Consequently, the colorbar limits differ significantly between the two analyses.

From a direct comparison with the active pixel map (Figure~\ref{fig:active_pixel}), we note that the map in  Figure~\ref{fig:std_combined}  clearly indicates that higher values of $\sigma_{\beta}$ strongly correlate with heavily masked patches.
This localized increase in variance is expected and it is due to the mask geometry, as  pseudo-$C_\ell$ estimators  suffer of highly irregular masks and low active-pixel fractions, with a severe reduction of  the local effective sky fraction ($f_{\mathrm{sky}}$). This reduction inevitably inflates the sample variance and complicates the mode-coupling inversion. However, it is crucial to emphasize that while the statistical uncertainty grows in these heavily masked regions, the estimator remains strictly unbiased.

\subsection{Estimates for forthcoming CMB probes}

\begin{figure}[htbp]
	\centering
	\includegraphics[width=0.6\linewidth]{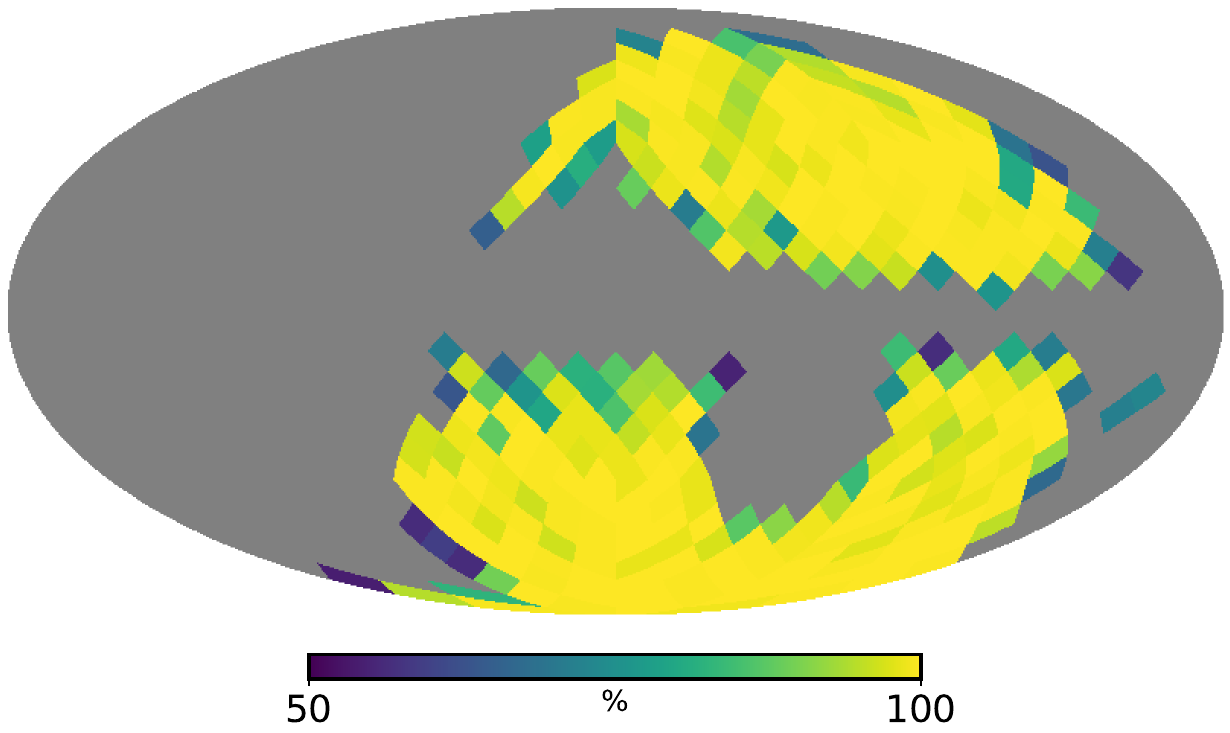}
	\caption{Percentage of active pixels for each patch. The gray patches have a fraction of active pixels  lower than 50\%.}
	\label{fig:active_pixel_sat}
\end{figure}

At this stage of the analysis, we present results for a prospective space-borne satellite mission and for a ground-based experiment. For the latter, we adopt an observational footprint that is realistically achievable for an instrument located in the Southern Hemisphere. 
Figure~\ref{fig:active_pixel_sat} illustrates the fraction of unmasked pixels per patch resulting from this masking procedure.

As shown by the comparison with Figure~\ref{fig:active_pixel}, the mask application lowers the active pixel fraction primarily along the boundaries, where the observed footprint has a sharp edge rather than the smooth Galactic cut used in
the main Planck-mask analysis. Patches straddling this sharp boundary have a highly irregular internal mask geometry, for which the analytic
pseudo-$C_\ell$ covariance approximation of section \ref{sec:pipeline} is less reliable. For
this reason, we adopt a stricter 90\% patch-retention threshold than the
50\% used for the  full-sky analysis, restricting the forecast to patches
whose coverage is close to uniform.

\begin{figure}[htbp]
    \centering
    
    \includegraphics[width=0.6\linewidth]{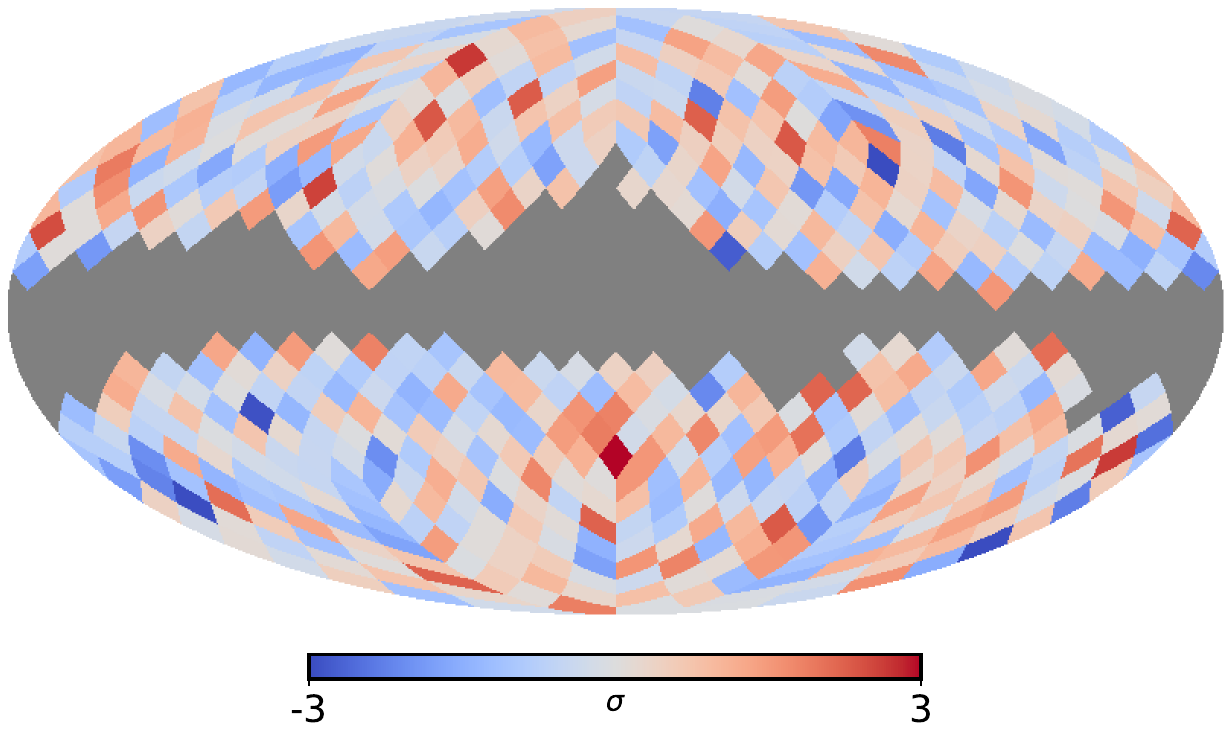}
    
    \vspace{0.4cm} 
    
    \includegraphics[width=0.6\linewidth]{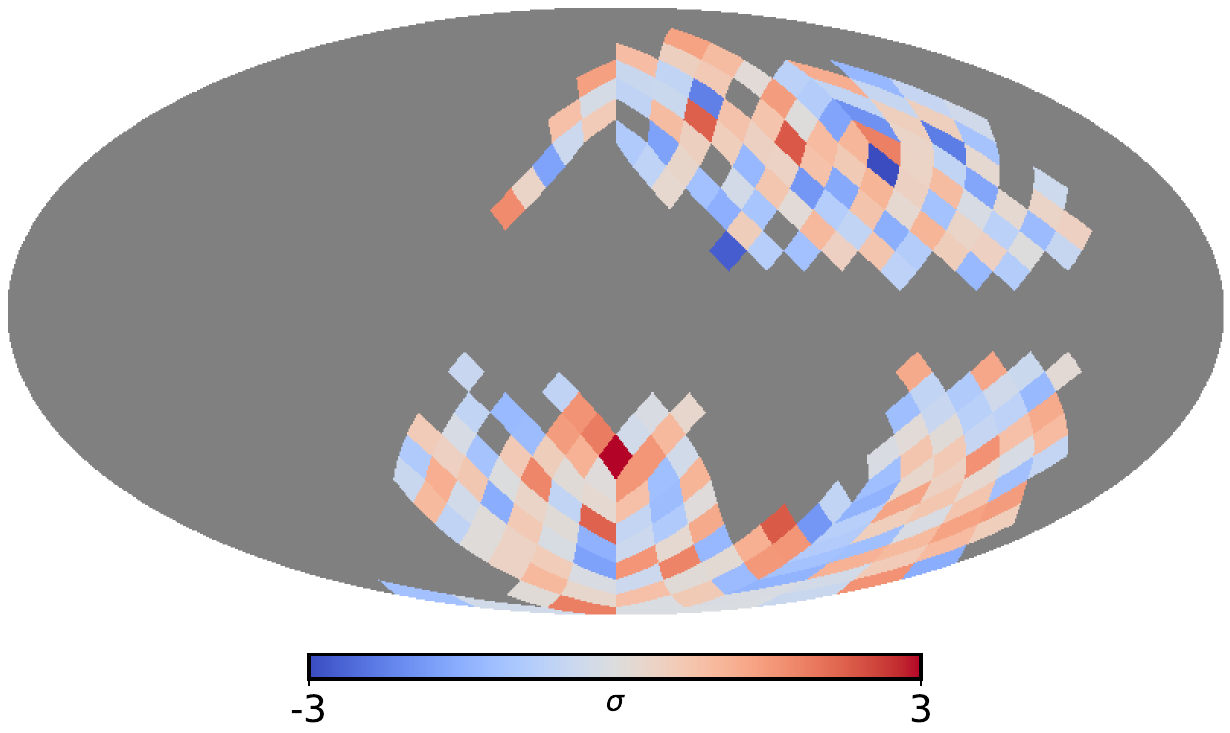}
    
    \caption{Null-test significance map of the recovered birefringence angle for space- (top) and ground- based (bottom) experiments. For each patch, the map displays the ratio between the mean recovered angle and its standard error, $\bar{\beta} / (\sigma / \sqrt{N_{\mathrm{sim}}})$, where $\sigma$ is the standard deviation evaluated over the $N_{\mathrm{sim}} = 1000$ simulated realizations. The lack of spatial structures confirms the unbiased nature of the pipeline.}
    \label{fig:residual_map}
\end{figure}

To quantitatively demonstrate the absence of systematic structures or spurious artifacts induced by the pipeline, we evaluate the local statistical significance of the extracted signal through a null-test. Specifically, for each patch, we compute the ratio between the mean recovered angle and its standard error, $\bar{\beta} / (\sigma / \sqrt{N_{\mathrm{sim}}})$, where $\sigma$ is the standard deviation evaluated over the $N_{\mathrm{sim}} = 1000$ simulated realizations. As demonstrated by the null-test significance map in Figure~\ref{fig:residual_map}, these values exhibit purely statistical fluctuations around zero. The lack of any spatially coherent clustering confirms that the masking and the inversion procedure do not inject any artificial parity-violating signal.

\begin{figure}[htbp]
    \centering
    
    \begin{minipage}[b]{0.48\textwidth}
        \centering
        \includegraphics[width=\textwidth]{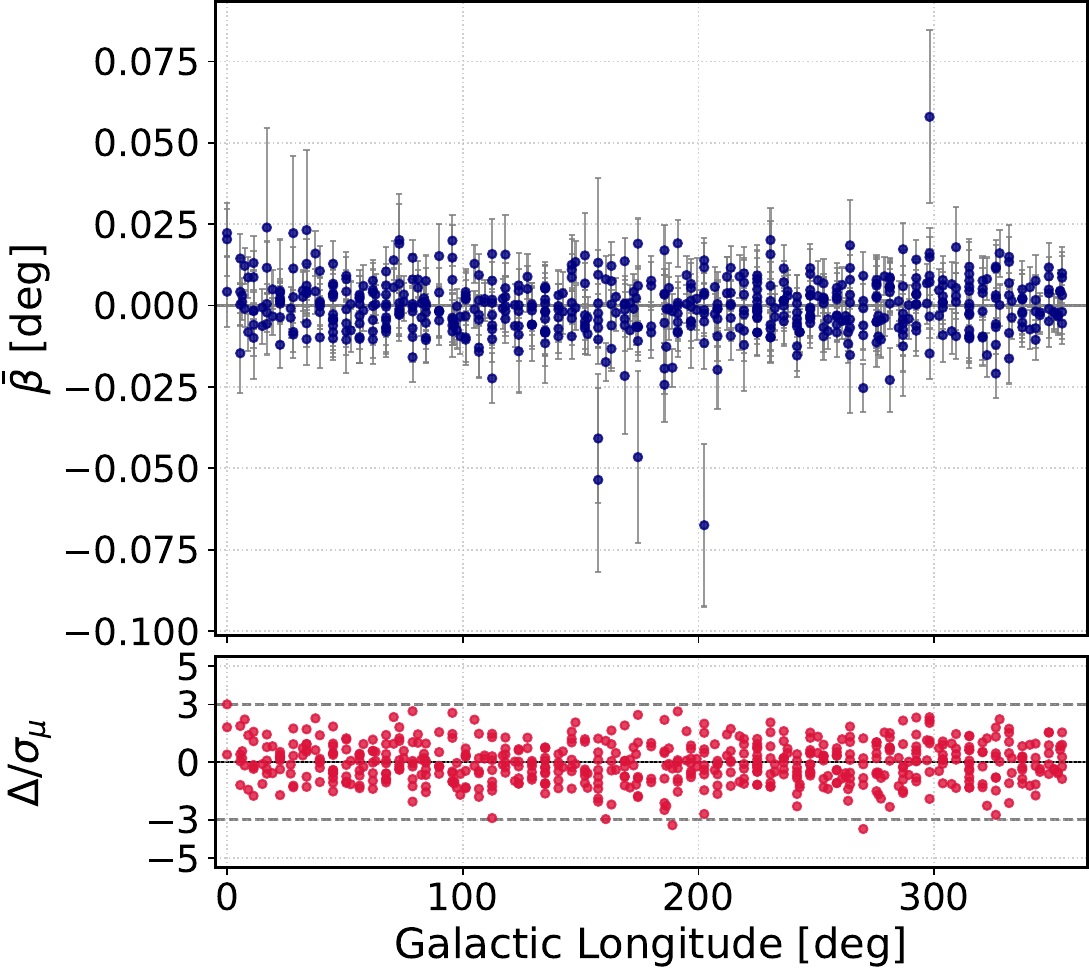}
    \end{minipage}
    \hfill
    \begin{minipage}[b]{0.48\textwidth}
        \centering
        \includegraphics[width=\textwidth]{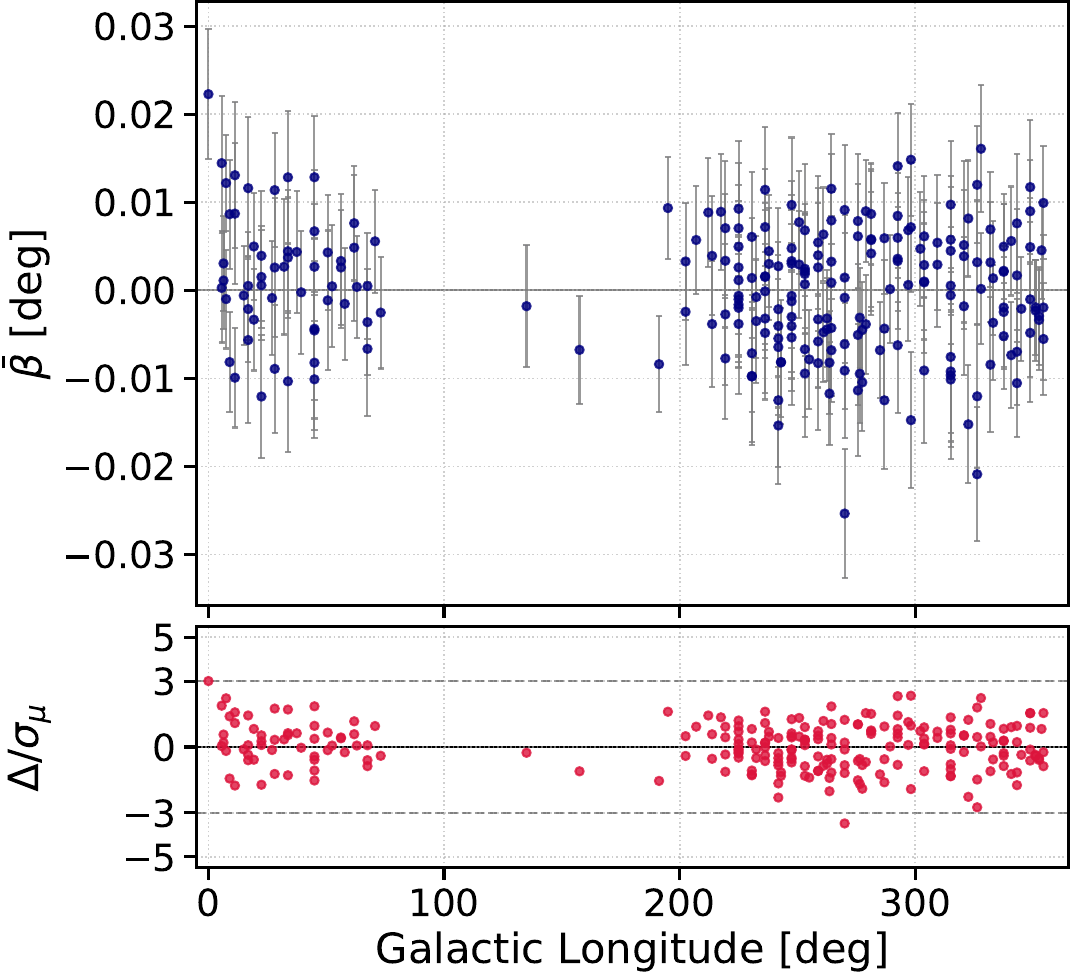}
    \end{minipage}

    \vspace{0.4cm} 
    \caption{Validation plot for space-satellite   (left) and SAT ground-based (right) simulations . Top panel: Mean value of the recovered angle $\bar{\beta}$ as a function of Galactic longitude. Bottom panel: Deviation in units of the standard error, $\Delta / \sigma_\mu$. Here, $\Delta$ represents the difference between the extracted angle and the expected null signal ($\bar{\beta} - 0$), and $\sigma_\mu$ is the standard error of the mean. The $\pm 3\sigma$ confidence intervals are marked by dashed lines.}
    \label{fig:residual_plot}
\end{figure}

This robustness test is further corroborated by Figure~\ref{fig:residual_plot}, which displays the mean recovered angle $\bar{\beta}$ as a function of Galactic longitude. To quantify the statistical significance of the extracted signal, the lower panel reports the normalized deviation from the expected null signal, denoted as $\Delta / \sigma_\mu$. In this context, $\Delta = \bar{\beta} - \beta_{\mathrm{model}}$ represents the absolute deviation from the theoretical expectation ($\beta_{\mathrm{model}} = 0$), while $\sigma_\mu = \sigma / \sqrt{N_{\mathrm{sim}}}$ is the standard error of the mean. As evident from the plot, these values are completely bounded within the $\pm 3\sigma$ statistical limits and do not exhibit any systematic dependence on longitude. This confirms that the simulated birefringence maps are perfectly compatible with a null signal across the entire sky.

\begin{figure}[htbp]
    \centering
    
    \begin{minipage}[b]{0.48\textwidth}
        \centering
        \includegraphics[width=\textwidth]{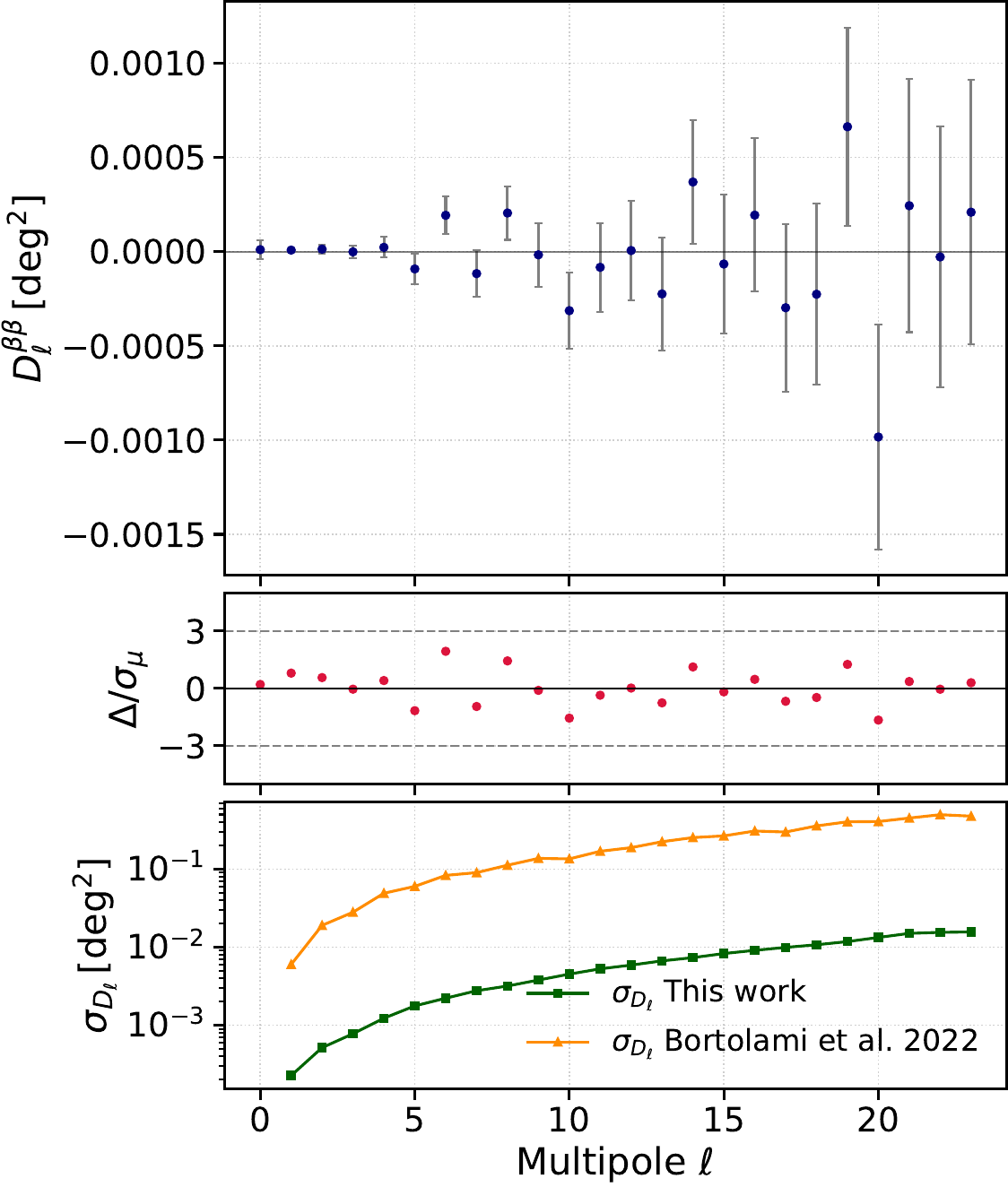}
    \end{minipage}
    \hfill
    \begin{minipage}[b]{0.48\textwidth}
        \centering
        \includegraphics[width=\textwidth]{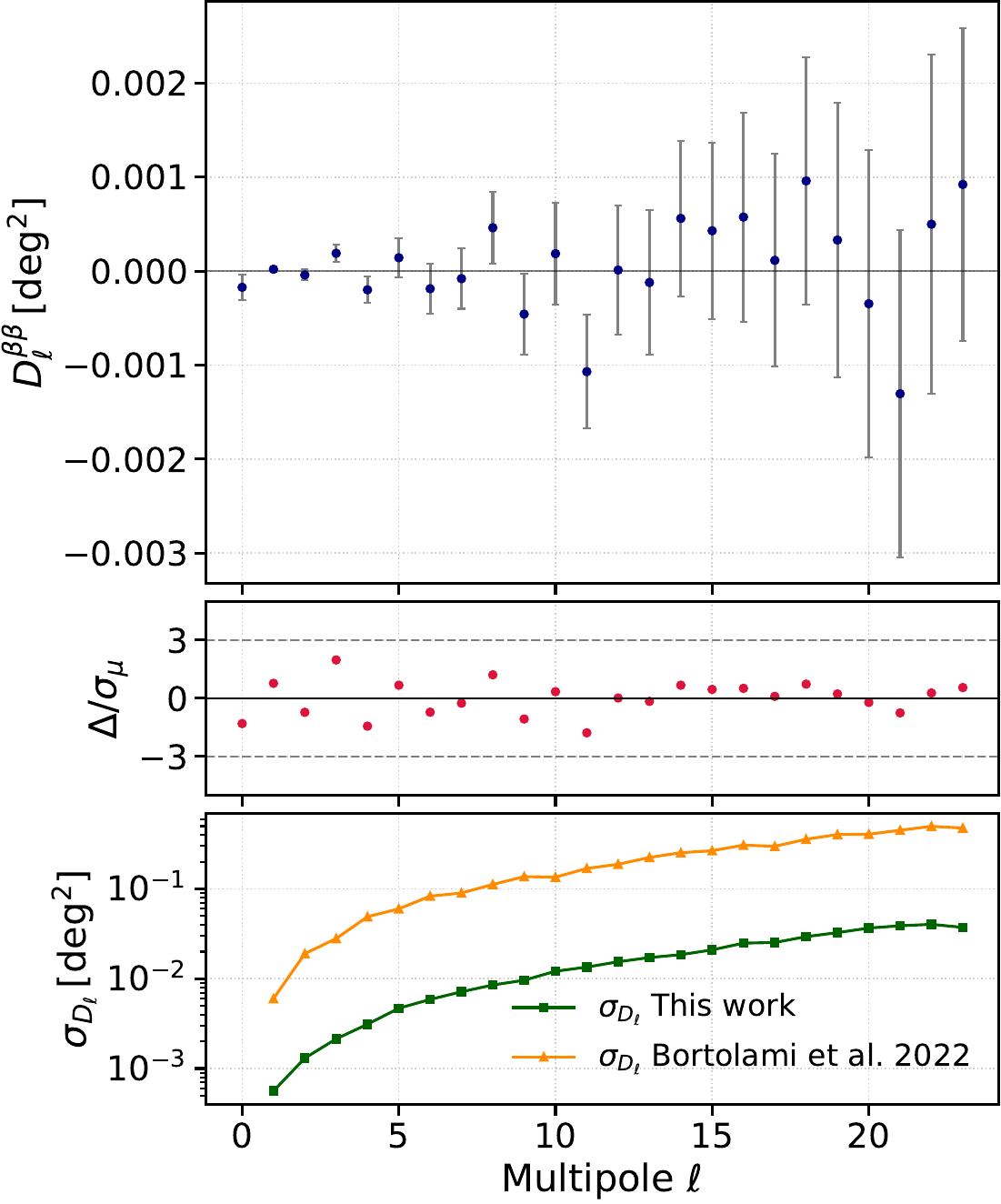}
    \end{minipage}

    \vspace{0.4cm} 
    \caption{Results for space-  (left) and SAT ground-based simulations (right). Top panel: Birefringence power spectrum validation. Middle panel: Deviation in units of the standard error $\Delta / \sigma_\mu$ with the $\pm 3\sigma$ confidence intervals. Bottom panel: Comparison between the standard deviation obtained in this work (green line) and the one reported for PR3 SMICA in~\cite{Bortolami2022}. The ratio between the two sigmas range from 26 to 40 for space-satellite case, and from 11 to 15 for  SAT ground-based case.}
    \label{fig:birefringence_plot}
\end{figure}

From the 1000 simulated birefringence maps, we compute the angular power spectrum of the birefringence angle utilizing \texttt{pymaster}. Following the same methodology used for the CMB spectra, for each experimental configuration (space-borne satellite and SAT ground-based)
we split its own 1000 birefringence angle maps into two independent subsets of 500 realizations each, sharing the same mask and
noise level. The CB angular power spectra are then evaluated in cross-mode
between the two subsets of the same configuration to ensure the removal of the statistical noise bias. 
The results are presented in Figure~\ref{fig:birefringence_plot}. The top and middle panels confirm that the recovered spectrum is fully consistent with a null signal, as expected. The bottom panel provides a direct comparison between the standard deviation achieved in this work and the limits set in~\cite{Bortolami2022}. It is evident how a nominal white noise level of $2\,\mu\text{K-arcmin}$ yields a drastically lower $\sigma$ compared to the noise levels characterizing the \textit{Planck} satellite data~\cite{Planck2}.

\begin{table}[htbp]
	\centering
	\renewcommand{\arraystretch}{1.2} 
	\begin{tabular}{c|cccc}
		\hline
		$\ell$ & $D_\ell^{\beta\beta} [\mathrm{deg}^2]$ & $\sigma_{D_\ell}^{\mathrm{This\,work}} [\mathrm{deg}^2]$ & $\sigma_{D_\ell}^{\mathrm{Bortolami}} [\mathrm{deg}^2]$ & $\sigma_{D_\ell}^{\mathrm{Bortolami}} / \sigma_{D_\ell}^{\mathrm{This\,work}}$ \\
		\hline
		1  & $ 8.13 \times 10^{-6}$ & $2.27 \times 10^{-4}$ & $0.006$ & $26.47$ \\
		2  & $ 1.31 \times 10^{-5}$ & $5.16 \times 10^{-4}$ & $0.019$ & $36.80$ \\
		3  & $-1.27 \times 10^{-6}$ & $7.78 \times 10^{-4}$ & $0.028$ & $36.00$ \\
		4  & $ 2.27 \times 10^{-5}$ & $1.23 \times 10^{-3}$ & $0.049$ & $39.84$ \\
		5  & $-9.19 \times 10^{-5}$ & $1.77 \times 10^{-3}$ & $0.060$ & $33.92$ \\
		6  & $ 1.93 \times 10^{-4}$ & $2.22 \times 10^{-3}$ & $0.083$ & $37.41$ \\
		7  & $-1.17 \times 10^{-4}$ & $2.77 \times 10^{-3}$ & $0.090$ & $32.54$ \\
		8  & $ 2.05 \times 10^{-4}$ & $3.18 \times 10^{-3}$ & $0.112$ & $35.23$ \\
		9  & $-1.70 \times 10^{-5}$ & $3.79 \times 10^{-3}$ & $0.137$ & $36.15$ \\
		10 & $-3.13 \times 10^{-4}$ & $4.52 \times 10^{-3}$ & $0.135$ & $29.90$ \\
		11 & $-8.32 \times 10^{-5}$ & $5.28 \times 10^{-3}$ & $0.169$ & $32.02$ \\
		12 & $ 5.88 \times 10^{-6}$ & $5.89 \times 10^{-3}$ & $0.188$ & $31.93$ \\
		13 & $-2.24 \times 10^{-4}$ & $6.67 \times 10^{-3}$ & $0.224$ & $33.57$ \\
		14 & $ 3.69 \times 10^{-4}$ & $7.34 \times 10^{-3}$ & $0.253$ & $34.46$ \\
		15 & $-6.58 \times 10^{-5}$ & $8.28 \times 10^{-3}$ & $0.266$ & $32.12$ \\
		16 & $ 1.94 \times 10^{-4}$ & $9.10 \times 10^{-3}$ & $0.307$ & $33.74$ \\
		17 & $-2.98 \times 10^{-4}$ & $9.93 \times 10^{-3}$ & $0.298$ & $30.01$ \\
		18 & $-2.26 \times 10^{-4}$ & $1.07 \times 10^{-2}$ & $0.357$ & $33.28$ \\
		19 & $ 6.63 \times 10^{-4}$ & $1.18 \times 10^{-2}$ & $0.403$ & $34.20$ \\
		20 & $-9.84 \times 10^{-4}$ & $1.33 \times 10^{-2}$ & $0.406$ & $30.51$ \\
		21 & $ 2.44 \times 10^{-4}$ & $1.50 \times 10^{-2}$ & $0.450$ & $29.99$ \\
		22 & $-2.81 \times 10^{-5}$ & $1.55 \times 10^{-2}$ & $0.497$ & $32.13$ \\
		23 & $ 2.09 \times 10^{-4}$ & $1.57 \times 10^{-2}$ & $0.476$ & $30.38$ \\
		\hline
	\end{tabular}
	\caption{Comparison between the standard deviation of the recovered spectra from this work for space-based and the reference values from PR3 SMICA Bortolami et al. ~\cite{Bortolami2022}. The final column highlights the improvement factor ($\sigma_{D_\ell}^{\mathrm{Bortolami}} / \sigma_{D_\ell}^{\mathrm{This\,work}}$).}
	\label{tab:sigma_comparison}
\end{table}

\begin{table}[htbp]
	\centering
	\renewcommand{\arraystretch}{1.2} 
	\begin{tabular}{c|cccc}
		\hline
		$\ell$ & $D_\ell^{\beta\beta} [\mathrm{deg}^2]$ & $\sigma_{D_\ell}^{\mathrm{This\,work}} [\mathrm{deg}^2]$ & $\sigma_{D_\ell}^{\mathrm{Bortolami}} [\mathrm{deg}^2]$ & $\sigma_{D_\ell}^{\mathrm{Bortolami}} / \sigma_{D_\ell}^{\mathrm{This\,work}}$ \\
		\hline
		1  & $ 1.95 \times 10^{-5}$ & $5.63 \times 10^{-4}$ & $0.006$ & $10.67$ \\
		2  & $-4.20 \times 10^{-5}$ & $1.31 \times 10^{-3}$ & $0.019$ & $14.56$ \\
		3  & $ 1.89 \times 10^{-4}$ & $2.14 \times 10^{-3}$ & $0.028$ & $13.10$ \\
		4  & $-2.00 \times 10^{-4}$ & $3.11 \times 10^{-3}$ & $0.049$ & $15.75$ \\
		5  & $ 1.42 \times 10^{-4}$ & $4.69 \times 10^{-3}$ & $0.060$ & $12.79$ \\
		6  & $-1.88 \times 10^{-4}$ & $5.88 \times 10^{-3}$ & $0.083$ & $14.12$ \\
		7  & $-8.06 \times 10^{-5}$ & $7.16 \times 10^{-3}$ & $0.090$ & $12.58$ \\
		8  & $ 4.61 \times 10^{-4}$ & $8.51 \times 10^{-3}$ & $0.112$ & $13.16$ \\
		9  & $-4.58 \times 10^{-4}$ & $9.62 \times 10^{-3}$ & $0.137$ & $14.24$ \\
		10 & $ 1.85 \times 10^{-4}$ & $1.21 \times 10^{-2}$ & $0.135$ & $11.14$ \\
		11 & $-1.07 \times 10^{-3}$ & $1.35 \times 10^{-2}$ & $0.169$ & $12.54$ \\
		12 & $ 1.09 \times 10^{-5}$ & $1.54 \times 10^{-2}$ & $0.188$ & $12.18$ \\
		13 & $-1.21 \times 10^{-4}$ & $1.72 \times 10^{-2}$ & $0.224$ & $13.02$ \\
		14 & $ 5.60 \times 10^{-4}$ & $1.85 \times 10^{-2}$ & $0.253$ & $13.67$ \\
		15 & $ 4.29 \times 10^{-4}$ & $2.09 \times 10^{-2}$ & $0.266$ & $12.70$ \\
		16 & $ 5.75 \times 10^{-4}$ & $2.49 \times 10^{-2}$ & $0.307$ & $12.33$ \\
		17 & $ 1.14 \times 10^{-4}$ & $2.53 \times 10^{-2}$ & $0.298$ & $11.78$ \\
		18 & $ 9.59 \times 10^{-4}$ & $2.94 \times 10^{-2}$ & $0.357$ & $12.14$ \\
		19 & $ 3.30 \times 10^{-4}$ & $3.26 \times 10^{-2}$ & $0.403$ & $12.36$ \\
		20 & $-3.48 \times 10^{-4}$ & $3.66 \times 10^{-2}$ & $0.406$ & $11.08$ \\
		21 & $-1.31 \times 10^{-3}$ & $3.88 \times 10^{-2}$ & $0.450$ & $11.58$ \\
		22 & $ 4.98 \times 10^{-4}$ & $4.03 \times 10^{-2}$ & $0.497$ & $12.33$ \\
		23 & $ 9.22 \times 10^{-4}$ & $3.73 \times 10^{-2}$ & $0.476$ & $12.78$ \\
		\hline
	\end{tabular}
    \caption{Comparison between the standard deviation of the recovered spectra from this work for SAT ground-based and the reference values from PR3 SMICA Bortolami et al.~\cite{Bortolami2022}. The final column highlights the improvement factor ($\sigma_{D_\ell}^{\mathrm{Bortolami}} / \sigma_{D_\ell}^{\mathrm{This\,work}}$).}
	\label{tab:sigma_comparison_sat}
\end{table}
The numerical results for each multipole are detailed in Tables~\ref{tab:sigma_comparison} and~\ref{tab:sigma_comparison_sat}. As highlighted in the final column, the statistical uncertainty $\sigma$ achieved in this forecasting setup is lower by a factor ranging from roughly 26 to nearly 40 for the satellite mission and from 11 to 15 for SAT ground-based compared to previous constraints. 

Once the anisotropic CB power spectrum is derived,  we set constraints in terms of the scale-invariant amplitude $A_{\rm CB}$, minimizing the following $\chi^2$, as in ~\cite{Bortolami2022}, :

\begin{equation}
    \chi^2(A_{\mathrm{CB}}) = \sum_{\ell,\ell'} \left[D_\ell^{\beta\beta,\mathrm{obs}} - A_{\mathrm{CB}}\right] (\mathrm{Cov}_{\ell\ell'})^{-1} \left[D_{\ell'}^{\beta\beta,\mathrm{obs}} - A_{\mathrm{CB}}\right], \label{eq:chi2_ACB}
\end{equation}

where $D_\ell^{\beta\beta,{\rm obs}}$ is the mean cross-spectrum, and ${\rm Cov}_{\ell\ell'}$ is its covariance matrix estimated from simulations.

\begin{figure}[htbp]
    \centering
    
    \begin{minipage}[b]{0.48\textwidth}
        \centering
        \includegraphics[width=\textwidth]{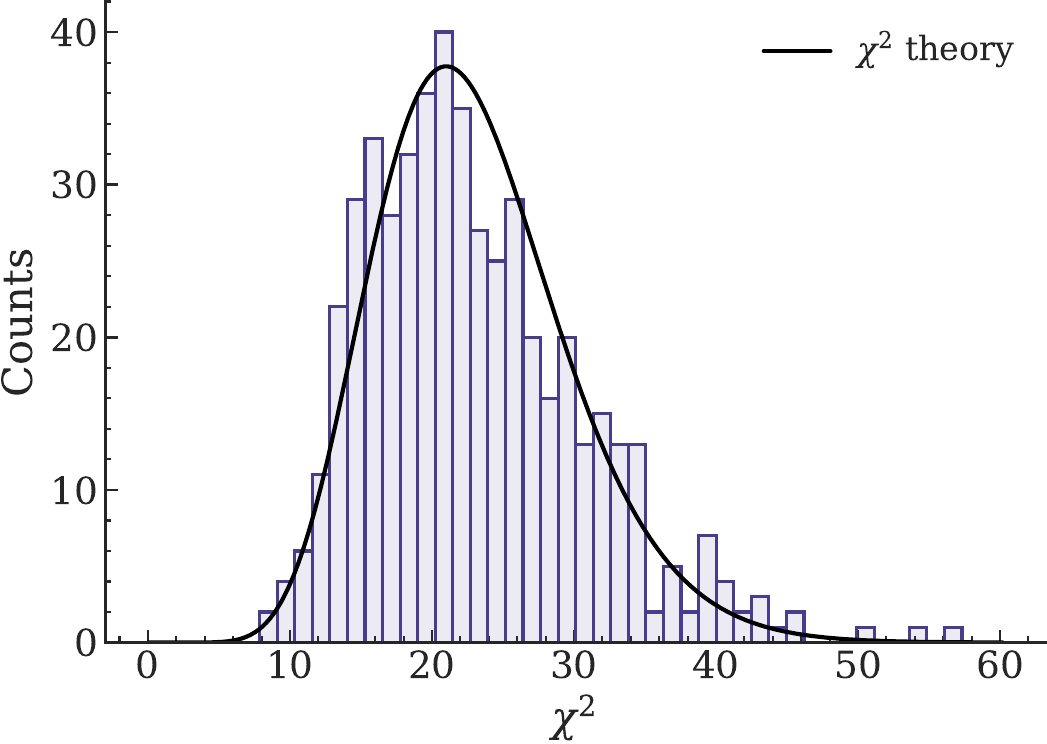}
    \end{minipage}
    \hfill
    \begin{minipage}[b]{0.48\textwidth}
        \centering
        \includegraphics[width=\textwidth]{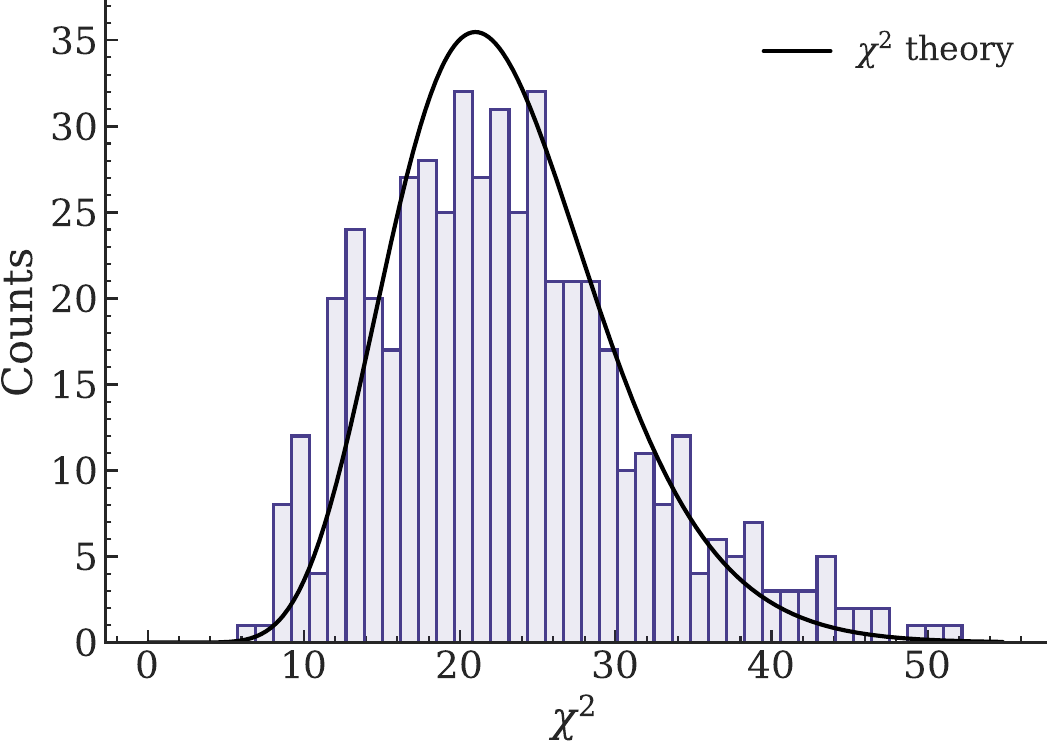}
    \end{minipage}

    \vspace{0.4cm} 
    \caption{Harmonic $\chi^2$ plots for space-(left) and SAT ground-based simulations (right). The solid black line represents the expected analytical $\chi^2$ probability density function with 23 degrees of freedom.}
    \label{fig:histo_acb}
\end{figure}
As a validation of the Gaussian likelihood of equation ~\ref{eq:chi2_ACB}, we evaluate the harmonic $\chi^2$ statistic at $A_{\rm CB}=0$ for each of the 500 independent realizations, and compare the resulting distribution to the theoretical $\chi^2_{n_\ell}$ expectation ($n_\ell=23$ multipole bins), shown in figure ~\ref{fig:histo_acb}. The good agreement confirms that the assumed Gaussianity and covariance estimate are adequate for this analysis.
\begin{table}[htbp]
	\centering
	\renewcommand{\arraystretch}{1.2} 
	\begin{tabular}{l|cccc}
		\hline
		 & $\hat{A}_{\mathrm{CB}} [\mathrm{deg}^2]$ & $\sigma_{A_{\mathrm{CB}}} [\mathrm{deg}^2]$ & $A_{\mathrm{CB}}$ 95\% C.L. $[\mathrm{deg}^2]$ & \makecell{Improvement factor \\ wrt \cite{Bortolami2022}} \\
		\hline
		Space-based   & $8.87 \times 10^{-6}$ & $1.74 \times 10^{-4}$ & $3.47 \times 10^{-4}$ & $\sim 20\times$ \\
		Ground-based  & $9.65 \times 10^{-6}$ & $3.59 \times 10^{-4}$ & $7.09 \times 10^{-4}$ & $\sim 9.9\times$ \\
		\hline
	\end{tabular}
    \caption{Results of the forecasted constraints on the scale-invariant amplitude of anisotropic CB, $A_{\mathrm{CB}}$. The table reports the best-fit values ($\hat{A}_{\mathrm{CB}}$), the $1\sigma$ statistical uncertainties ($\sigma_{A_{\mathrm{CB}}}$), and the 95\% C.L. upper limits for both the satellite and ground-based SAT configurations. The final column highlights the improvement factor of our upper limits compared to the constraints from~\cite{Bortolami2022}.}
    \label{tab:acb_results}
\end{table}
\begin{figure}[htbp]
    \centering
    
    \begin{minipage}[b]{0.48\textwidth}
        \centering
        \includegraphics[width=\textwidth]{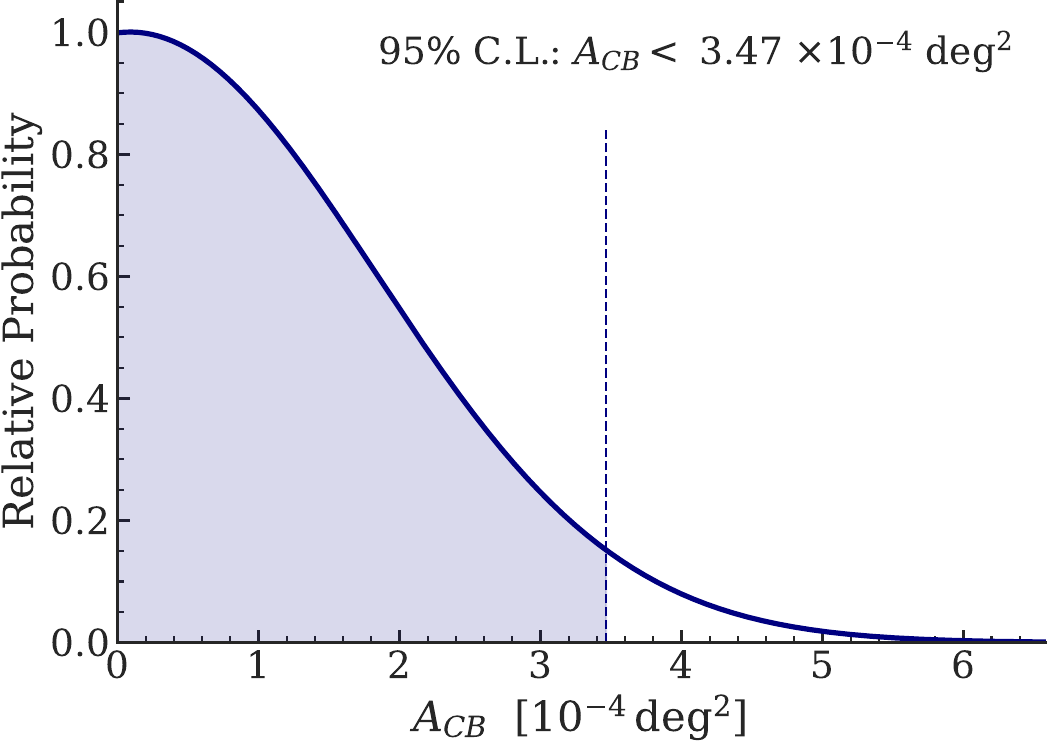}
    \end{minipage}
    \hfill
    \begin{minipage}[b]{0.48\textwidth}
        \centering
        \includegraphics[width=\textwidth]{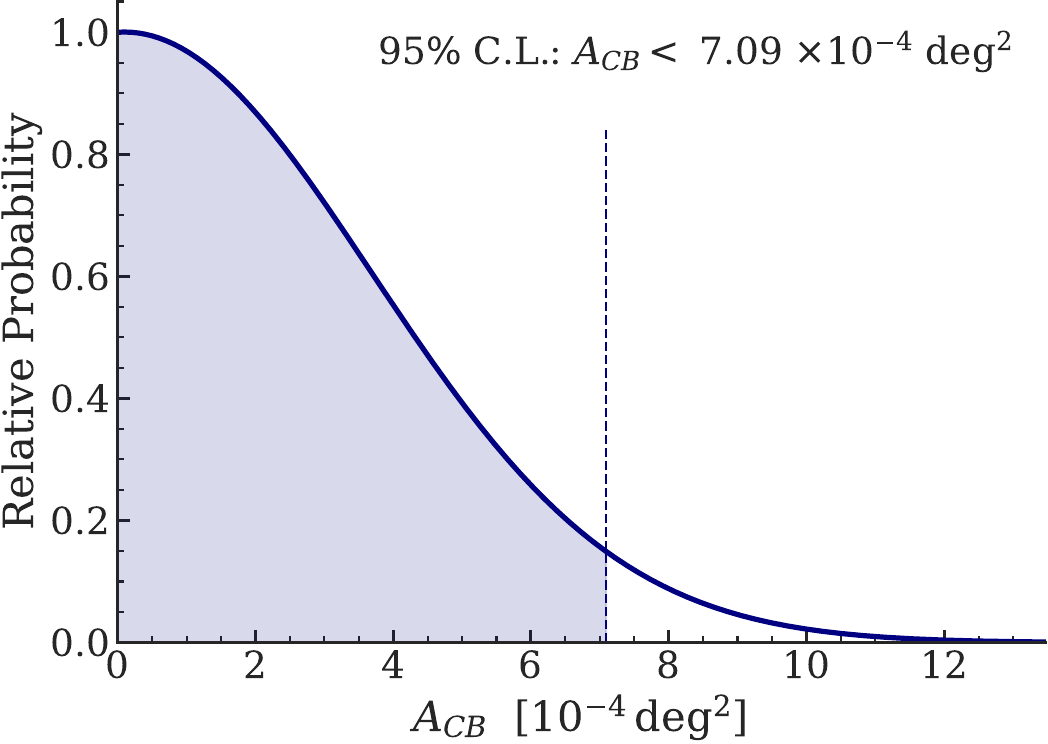}
    \end{minipage}

    \vspace{0.4cm} 
    \caption{Posterior distributions of the scale-invariant amplitude $A_{\mathrm{CB}}$ for space-based (left) and ground-based (right) experimental configurations. The vertical dashed lines indicate the 95\% C.L. upper limits.}
    \label{fig:posterior_acb}
\end{figure}
Figure~\ref{fig:posterior_acb} show the posterior distribution of the scale-invariant amplitude $A_{\mathrm{CB}}$ computed using equation~\ref{eq:chi2_ACB}. We find $A_{\rm CB} < 3.5\times10^{-4}\ {\rm deg}^2$ for space-based and $A_{\rm CB} < 7.1\times10^{-4}\ {\rm deg}^2$ for ground-based experiments at 95\% C.L., summarized in table \ref{tab:acb_results} together with the corresponding improvement factor over the results from Bortolami et al. ($A_{\mathrm{CB}}<0.007\,{\rm deg}^2$~\cite{Bortolami2022}).

Although  constraints in~\cite{Bortolami2022} are derived from real Planck data, encoding residual Galactic foregrounds, correlated noise, and instrumental systematics, we note that, unlike Planck, future experiments are explicitly designed to control residual instrumental systematics well below their statistical sensitivity ~\cite{LiteBIRD:2022cnt,SimonsObservatory:2018koc}. The purely statistical forecast presented here may therefore represent a closer approximation to the achievable sensitivity of these future experiments than the comparison with Planck alone would suggest, even though it remains an idealization in other respects.

\newpage
\section{Conclusions}
\label{sec:conclusion}

Cosmic birefringence, defined as the in-vacuo rotation of the polarization plane of the Cosmic Microwave Background (CMB), represents a fundamental probe to explore new physics and potential parity violations. While recent analyses of \textit{Planck} data have provided growing evidence for a non-null isotropic birefringence angle, there is currently no statistically significant detection of the anisotropic component of the signal. 

Therefore, the primary objective of this work is to forecast the statistical
uncertainties achievable by future CMB experiments, by quantifying the
expected error on the rotation angle anisotropies and its scale-dependent
variance under simplified, idealized simulations. To achieve this goal, a rigorous data analysis pipeline is developed and validated, where  1000 simulated CMB maps based on \textit{Planck} fiducial model are combined with independent sets of uncorrelated noise maps. In order to extract spatial anisotropies, the full-sky map is divided into local patches with an HEALPix resolution of Nside=8.

To ensure statistical reliability, the analysis is strictly restricted to patches containing at least 50\% active pixels, and apodized masks are applied to mitigate Galactic foreground contamination and the leakage between E and B modes. Within each patch, the pseudo-$C_\ell$ power spectra were computed using the \texttt{pymaster} algorithm. Subsequently, the D-estimator, a statistical framework employed under the assumption of local isotropy, is applied to minimize the $\chi^2$ and extract the local rotation angle $\beta$ in every single isolated region of the sky. The systematic iteration of this procedure across the entire set of simulations successfully allowed for the reconstruction of 1000 full-sky directional maps of the birefringence angle. 

The analysis of these maps confirmed that the developed algorithm is unbiased and does not introduce systematic structures or artifacts. Although the standard deviation of the extracted signal naturally increases in heavily masked regions near the Galactic plane due to the reduction of the effective sky fraction, the estimators remain unbiased.
From the reconstructed maps, the angular power spectrum of the anisotropic birefringence is derived by cross-correlating independent CB angle maps, effectively eliminating the noise bias. The final results of this work quantitatively demonstrate the unprecedented sensitivity of future space and ground-based missions. The statistical uncertainty obtained by simulating the ideal conditions of these future experiments is lower by a factor ranging from 26 to 40 for the space-based and from 11 to 15 for SAT ground-based experiments, compared to the current constraints derived from \textit{Planck} data using the same methodology. Fitting the reconstructed spectra with a scale-invariant amplitude model, we further forecast $A_{\rm CB} < 3.5\times10^{-4}\,{\rm deg}^2$ from space-based and $A_{\rm CB} < 7.1\times10^{-4}\,{\rm deg}^2$ from ground-based experiments, improving by a factor of about 20 and 10 compared to the aforementioned benchmark.

We remark that,  although the methodology is applied  on simplified simulated datasets, we aim at assessing its performance when applied on the forthcoming data sets.  In a future work, we plan to  include more  realistic instrumental effects,  correlated $(1/f)$ noise, which would affect the covariance matrices.
Moreover, we highlight that the present forecast  relying exclusively on the PR3 masks (Figure~\ref{fig:masks}) to exclude the most heavily contaminated regions but does not account for foreground emission. As demonstrated by  ~\cite{Diego-Palazuelos:2022cnh}, while conservative masking mitigates foreground effects, residual dust-induced $EB$ correlations can still induce a bias in the recovered birefringence signal if not explicitly modeled.

As a natural extension of this work, we plan to carry out a dedicated robustness analysis in the same spirit as~\cite{Diego-Palazuelos:2022cnh}. By propagating realistic Galactic dust templates and representative instrumental systematics (e.g., polarization-angle miscalibration, differential beam effects, and bandpass mismatch) through our patch-based D-estimator pipeline, we will directly quantify their impact on the recovered anisotropic birefringence power spectrum and the resulting upper limits.

\acknowledgments
The authors are very thankful to P. Diego-Palazuelos for carefully reading and commenting on the manuscript.   
We acknowledge the use of the \texttt{scipy~\cite{2020SciPy-NMeth} , healpy~\cite{Zonca2019}~\cite{Gorski2005}, numpy~\cite{harris2020array}, pymaster~\cite{Alonso2019}} software packages. This research used resources of the National Energy Research Scientific Computing Center (NERSC), a Department of Energy User Facility (project mp107d).


 \bibliographystyle{JHEP}
 \bibliography{biblio.bib}

@article{Planck2018_CosmologicalParameters,
    author        = {Planck Collaboration},
    title         = "{Planck 2018 results. VI. Cosmological parameters}",
    journal       = {Astronomy \& Astrophysics},
    volume        = {641},
    pages         = {A6},
    year          = {2020},
    doi           = {10.1051/0004-6361/201833910},
    archivePrefix = {arXiv},
    eprint        = {1807.06209},
    primaryClass  = {astro-ph.CO}
}

@article{Carroll1990_Lorentz,
    author        = {Carroll, Sean M. and Field, George B. and Jackiw, Roman},
    title         = "{Limits on a Lorentz- and parity-violating modification of electrodynamics}",
    journal       = {Physical Review D},
    volume        = {41},
    issue         = {4},
    pages         = {1231},
    year          = {1990},
    doi           = {10.1103/PhysRevD.41.1231},
    publisher     = {American Physical Society}
}

@article{Marsh2016_AxionCosmology,
    author        = {Marsh, David J. E.},
    title         = "{Axion cosmology}",
    journal       = {Physics Reports},
    volume        = {643},
    pages         = {1--79},
    year          = {2016},
    doi           = {10.1016/j.physrep.2016.06.005},
    archivePrefix = {arXiv},
    eprint        = {1510.07633},
    primaryClass  = {astro-ph.CO}
}

@article{Lee1956,
  title = {Question of Parity Conservation in Weak Interactions},
  volume = {104},
  ISSN = {0031-899X},
  url = {http://dx.doi.org/10.1103/PhysRev.104.254},
  DOI = {10.1103/physrev.104.254},
  number = {1},
  journal = {Physical Review},
  publisher = {American Physical Society (APS)},
  author = {Lee,  T. D. and Yang,  C. N.},
  year = {1956},
  month = oct,
  pages = {254–258}
}

@article{Wu1957,
  title = {Experimental Test of Parity Conservation in Beta Decay},
  volume = {105},
  ISSN = {0031-899X},
  url = {http://dx.doi.org/10.1103/PhysRev.105.1413},
  DOI = {10.1103/physrev.105.1413},
  number = {4},
  journal = {Physical Review},
  publisher = {American Physical Society (APS)},
  author = {Wu,  C. S. and Ambler,  E. and Hayward,  R. W. and Hoppes,  D. D. and Hudson,  R. P.},
  year = {1957},
  month = feb,
  pages = {1413–1415}
}

@article{Peccei1977,
  title = {CP Conservation in the Presence of Pseudoparticles},
  volume = {38},
  ISSN = {0031-9007},
  url = {http://dx.doi.org/10.1103/PhysRevLett.38.1440},
  DOI = {10.1103/physrevlett.38.1440},
  number = {25},
  journal = {Physical Review Letters},
  publisher = {American Physical Society (APS)},
  author = {Peccei,  R. D. and Quinn,  Helen R.},
  year = {1977},
  month = jun,
  pages = {1440–1443}
}

@article{Weinberg1978,
  title = {A New Light Boson?},
  volume = {40},
  ISSN = {0031-9007},
  url = {http://dx.doi.org/10.1103/PhysRevLett.40.223},
  DOI = {10.1103/physrevlett.40.223},
  number = {4},
  journal = {Physical Review Letters},
  publisher = {American Physical Society (APS)},
  author = {Weinberg,  Steven},
  year = {1978},
  month = jan,
  pages = {223–226}
}

@article{Lue1999,
  title = {Cosmological Signature of New Parity-Violating Interactions},
  volume = {83},
  ISSN = {1079-7114},
  url = {http://dx.doi.org/10.1103/PhysRevLett.83.1506},
  DOI = {10.1103/physrevlett.83.1506},
  number = {8},
  journal = {Physical Review Letters},
  publisher = {American Physical Society (APS)},
  author = {Lue,  Arthur and Wang,  Limin and Kamionkowski,  Marc},
  year = {1999},
  month = aug,
  pages = {1506–1509}
}

@article{Finelli2009,
  title = {Rotation of linear polarization plane and circular polarization from cosmological pseudoscalar fields},
  volume = {79},
  ISSN = {1550-2368},
  url = {http://dx.doi.org/10.1103/PhysRevD.79.063002},
  DOI = {10.1103/physrevd.79.063002},
  number = {6},
  journal = {Physical Review D},
  publisher = {American Physical Society (APS)},
  author = {Finelli,  Fabio and Galaverni,  Matteo},
  year = {2009},
  month = mar 
}

@article{Komatsu2022,
  title = {New physics from the polarized light of the cosmic microwave background},
  volume = {4},
  ISSN = {2522-5820},
  url = {http://dx.doi.org/10.1038/s42254-022-00452-4},
  DOI = {10.1038/s42254-022-00452-4},
  number = {7},
  journal = {Nature Reviews Physics},
  publisher = {Springer Science and Business Media LLC},
  author = {Komatsu,  Eiichiro},
  year = {2022},
  month = may,
  pages = {452–469}
}

@article{Bortolami2022,
  title = {Planck constraints on cross-correlations between anisotropic cosmic birefringence and CMB polarization},
  volume = {2022},
  ISSN = {1475-7516},
  url = {http://dx.doi.org/10.1088/1475-7516/2022/09/075},
  DOI = {10.1088/1475-7516/2022/09/075},
  number = {09},
  journal = {Journal of Cosmology and Astroparticle Physics},
  publisher = {IOP Publishing},
  author = {Bortolami,  M. and Billi,  M. and Gruppuso,  A. and Natoli,  P. and Pagano,  L.},
  year = {2022},
  month = sep,
  pages = {075}
}

@article{Gorski2005,
  title = {HEALPix: A Framework for High‐Resolution Discretization and Fast Analysis of Data Distributed on the Sphere},
  volume = {622},
  ISSN = {1538-4357},
  url = {http://dx.doi.org/10.1086/427976},
  DOI = {10.1086/427976},
  number = {2},
  journal = {The Astrophysical Journal},
  publisher = {American Astronomical Society},
  author = {Gorski,  K. M. and Hivon,  E. and Banday,  A. J. and Wandelt,  B. D. and Hansen,  F. K. and Reinecke,  M. and Bartelmann,  M.},
  year = {2005},
  month = apr,
  pages = {759–771}
}

@article{Alonso2019,
  title = {A unified pseudo-Cl framework},
  volume = {484},
  ISSN = {1365-2966},
  url = {http://dx.doi.org/10.1093/mnras/stz093},
  DOI = {10.1093/mnras/stz093},
  number = {3},
  journal = {Monthly Notices of the Royal Astronomical Society},
  publisher = {Oxford University Press (OUP)},
  author = {Alonso,  David and Sanchez,  Javier and Slosar,  Anže},
  year = {2019},
  month = jan,
  pages = {4127–4151}
}

@article{Gruppuso2020,
  title = {Planck 2018 constraints on anisotropic birefringence and its cross-correlation with CMB anisotropy},
  volume = {2020},
  ISSN = {1475-7516},
  url = {http://dx.doi.org/10.1088/1475-7516/2020/11/066},
  DOI = {10.1088/1475-7516/2020/11/066},
  number = {11},
  journal = {Journal of Cosmology and Astroparticle Physics},
  publisher = {IOP Publishing},
  author = {Gruppuso,  A. and Molinari,  D. and Natoli,  P. and Pagano,  L.},
  year = {2020},
  month = nov,
  pages = {066–066}
}

@article{l50Garcia2019,
  title = {Disconnected pseudo-Cl covariances for projected large-scale structure data},
  volume = {2019},
  ISSN = {1475-7516},
  url = {http://dx.doi.org/10.1088/1475-7516/2019/11/043},
  DOI = {10.1088/1475-7516/2019/11/043},
  number = {11},
  journal = {Journal of Cosmology and Astroparticle Physics},
  publisher = {IOP Publishing},
  author = {García-García,  Carlos and Alonso,  David and Bellini,  Emilio},
  year = {2019},
  month = nov,
  pages = {043–043}
}

@article{l50Planck,
  title = {Planck2018 results: V. CMB power spectra and likelihoods},
  volume = {641},
  ISSN = {1432-0746},
  url = {http://dx.doi.org/10.1051/0004-6361/201936386},
  DOI = {10.1051/0004-6361/201936386},
  journal = {Astronomy \& Astrophysics},
  publisher = {EDP Sciences},
  author = {Planck Collaboration},
  year = {2020},
  month = sep,
  pages = {A5}
}

@article{l50Couchot2017,
  title = {Cosmology with the cosmic microwave background temperature-polarization correlation},
  volume = {602},
  ISSN = {1432-0746},
  url = {http://dx.doi.org/10.1051/0004-6361/201629815},
  DOI = {10.1051/0004-6361/201629815},
  journal = {Astronomy \& Astrophysics},
  publisher = {EDP Sciences},
  author = {Couchot,  F. and Henrot-Versillé,  S. and Perdereau,  O. and Plaszczynski,  S. and Rouillé d’Orfeuil,  B. and Spinelli,  M. and Tristram,  M.},
  year = {2017},
  month = jun,
  pages = {A41}
}

@article{2020Planck3,
  title = {Planck 2018 results: III. High Frequency Instrument data processing and frequency maps},
  volume = {641},
  ISSN = {1432-0746},
  url = {http://dx.doi.org/10.1051/0004-6361/201832909},
  DOI = {10.1051/0004-6361/201832909},
  journal = {Astronomy \& Astrophysics},
  publisher = {EDP Sciences},
  author = {Planck Collaboration},
  year = {2020},
  month = sep,
  pages = {A3}
}

@article{2020Planck7,
  title = {Planck2018 results: VII. Isotropy and statistics of the CMB},
  volume = {641},
  ISSN = {1432-0746},
  url = {http://dx.doi.org/10.1051/0004-6361/201935201},
  DOI = {10.1051/0004-6361/201935201},
  journal = {Astronomy \& Astrophysics},
  publisher = {EDP Sciences},
  author = {Planck Collaboration},
  year = {2020},
  month = sep,
  pages = {A7}
}

@article{Planck2,
  title = {Planck2018 results: I. Overview and the cosmological legacy of Planck},
  volume = {641},
  ISSN = {1432-0746},
  url = {http://dx.doi.org/10.1051/0004-6361/201833880},
  DOI = {10.1051/0004-6361/201833880},
  journal = {Astronomy \& Astrophysics},
  publisher = {EDP Sciences},
  author = {Planck Collaboration},
  year = {2020},
  month = sep,
  pages = {A1}
}

@article{Minami:2020fin,
    author = "Minami, Yuto and Komatsu, Eiichiro",
    title = "{Simultaneous determination of the cosmic birefringence and miscalibrated polarization angles II: Including cross frequency spectra}",
    eprint = "2006.15982",
    archivePrefix = "arXiv",
    primaryClass = "astro-ph.CO",
    doi = "10.1093/ptep/ptaa130",
    journal = "PTEP",
    volume = "2020",
    number = "10",
    pages = "103E02",
    year = "2020"
}

@article{Diego-Palazuelos:2022cnh,
    author = "Diego-Palazuelos, P. and others",
    title = "{Robustness of cosmic birefringence measurement against Galactic foreground emission and instrumental systematics}",
    eprint = "2210.07655",
    archivePrefix = "arXiv",
    primaryClass = "astro-ph.CO",
    doi = "10.1088/1475-7516/2023/01/044",
    journal = "JCAP",
    volume = "01",
    pages = "044",
    year = "2023"
}

@article{LiteBIRD:2022cnt,
    author = "Allys, E. and others",
    collaboration = "LiteBIRD",
    title = "{Probing Cosmic Inflation with the LiteBIRD Cosmic Microwave Background Polarization Survey}",
    eprint = "2202.02773",
    archivePrefix = "arXiv",
    primaryClass = "astro-ph.IM",
    doi = "10.1093/ptep/ptac150",
    journal = "PTEP",
    volume = "2023",
    number = "4",
    pages = "042F01",
    year = "2023"
}

@article{SimonsObservatory:2018koc,
    author = "Ade, Peter and others",
    collaboration = "Simons Observatory",
    title = "{The Simons Observatory: Science goals and forecasts}",
    eprint = "1808.07445",
    archivePrefix = "arXiv",
    primaryClass = "astro-ph.CO",
    doi = "10.1088/1475-7516/2019/02/056",
    journal = "JCAP",
    volume = "02",
    pages = "056",
    year = "2019"
}

@article{Bicep/keck,
  title = {Improved Constraints on Primordial Gravitational Waves using 
Planck
,  WMAP,  and BICEP/
Keck
 Observations through the 2018 Observing Season},
  volume = {127},
  ISSN = {1079-7114},
  url = {http://dx.doi.org/10.1103/PhysRevLett.127.151301},
  DOI = {10.1103/physrevlett.127.151301},
  number = {15},
  journal = {Physical Review Letters},
  publisher = {American Physical Society (APS)},
  author = {Ade, P.A.R. and others },
  year = {2021},
  month = oct 
}

@article{PhysRevLett.102.161302,
  title = {Parity Violation Constraints Using Cosmic Microwave Background Polarization Spectra from 2006 and 2007 Observations by the QUaD Polarimeter},
  author = {Wu, E. Y. S. and Ade, P. and Bock, J. and Bowden, M. and Brown, M. L. and Cahill, G. and Castro, P. G. and Church, S. and Culverhouse, T. and Friedman, R. B. and Ganga, K. and Gear, W. K. and Gupta, S. and Hinderks, J. and Kovac, J. and Lange, A. E. and Leitch, E. and Melhuish, S. J. and Memari, Y. and Murphy, J. A. and Orlando, A. and Piccirillo, L. and Pryke, C. and Rajguru, N. and Rusholme, B. and Schwarz, R. and O'Sullivan, C. and Taylor, A. N. and Thompson, K. L. and Turner, A. H. and Zemcov, M.},
  collaboration = {QUaD Collaboration},
  journal = {Phys. Rev. Lett.},
  volume = {102},
  issue = {16},
  pages = {161302},
  numpages = {4},
  year = {2009},
  month = {Apr},
  publisher = {American Physical Society},
  doi = {10.1103/PhysRevLett.102.161302},
  url = {https://link.aps.org/doi/10.1103/PhysRevLett.102.161302}
}

@article{hcsm-39qw,
  title = {Cosmic birefringence from a joint analysis of ACT and Planck},
  author = {Eskilt, Johannes R.},
  journal = {Phys. Rev. D},
  volume = {114},
  issue = {4},
  pages = {043553},
  numpages = {8},
  year = {2026},
  month = {Aug},
  publisher = {American Physical Society},
  doi = {10.1103/hcsm-39qw},
  url = {https://link.aps.org/doi/10.1103/hcsm-39qw}
}

@article{2020SciPy-NMeth,
  author  = {Virtanen, Pauli and Gommers, Ralf and Oliphant, Travis E. and
            Haberland, Matt and Reddy, Tyler and Cournapeau, David and
            Burovski, Evgeni and Peterson, Pearu and Weckesser, Warren and
            Bright, Jonathan and {van der Walt}, St{\'e}fan J. and
            Brett, Matthew and Wilson, Joshua and Millman, K. Jarrod and
            Mayorov, Nikolay and Nelson, Andrew R. J. and Jones, Eric and
            Kern, Robert and Larson, Eric and Carey, C J and
            Polat, {\.I}lhan and Feng, Yu and Moore, Eric W. and
            {VanderPlas}, Jake and Laxalde, Denis and Perktold, Josef and
            Cimrman, Robert and Henriksen, Ian and Quintero, E. A. and
            Harris, Charles R. and Archibald, Anne M. and
            Ribeiro, Ant{\^o}nio H. and Pedregosa, Fabian and
            {van Mulbregt}, Paul and {SciPy 1.0 Contributors}},
  title   = {{{SciPy} 1.0: Fundamental Algorithms for Scientific
            Computing in Python}},
  journal = {Nature Methods},
  year    = {2020},
  volume  = {17},
  pages   = {261--272},
  adsurl  = {https://rdcu.be/b08Wh},
  doi     = {10.1038/s41592-019-0686-2},
}

@article{Zonca2019,
  doi = {10.21105/joss.01298},
  url = {https://doi.org/10.21105/joss.01298},
  year = {2019},
  month = mar,
  publisher = {The Open Journal},
  volume = {4},
  number = {35},
  pages = {1298},
  author = {Andrea Zonca and Leo Singer and Daniel Lenz and Martin Reinecke and Cyrille Rosset and Eric Hivon and Krzysztof Gorski},
  title = {healpy: equal area pixelization and spherical harmonics transforms for data on the sphere in Python},
  journal = {Journal of Open Source Software}
}

@Article{         harris2020array,
 title         = {Array programming with {NumPy}},
 author        = {Charles R. Harris and K. Jarrod Millman and St{\'{e}}fan J.
                 van der Walt and Ralf Gommers and Pauli Virtanen and David
                 Cournapeau and Eric Wieser and Julian Taylor and Sebastian
                 Berg and Nathaniel J. Smith and Robert Kern and Matti Picus
                 and Stephan Hoyer and Marten H. van Kerkwijk and Matthew
                 Brett and Allan Haldane and Jaime Fern{\'{a}}ndez del
                 R{\'{i}}o and Mark Wiebe and Pearu Peterson and Pierre
                 G{\'{e}}rard-Marchant and Kevin Sheppard and Tyler Reddy and
                 Warren Weckesser and Hameer Abbasi and Christoph Gohlke and
                 Travis E. Oliphant},
 year          = {2020},
 month         = sep,
 journal       = {Nature},
 volume        = {585},
 number        = {7825},
 pages         = {357--362},
 doi           = {10.1038/s41586-020-2649-2},
 publisher     = {Springer Science and Business Media {LLC}},
 url           = {https://doi.org/10.1038/s41586-020-2649-2}
}

@article{RevModPhys.82.557,
  title = {Axions and the strong $CP$ problem},
  author = {Kim, Jihn E. and Carosi, Gianpaolo},
  journal = {Rev. Mod. Phys.},
  volume = {82},
  issue = {1},
  pages = {557--601},
  numpages = {0},
  year = {2010},
  month = {Mar},
  publisher = {American Physical Society},
  doi = {10.1103/RevModPhys.82.557},
  url = {https://link.aps.org/doi/10.1103/RevModPhys.82.557}
}

@article{PhysRevLett.128.091302,
  title = {Cosmic Birefringence from the Planck Data Release 4},
  author = {Diego-Palazuelos, P. and Eskilt, J. R. and Minami, Y. and Tristram, M. and Sullivan, R. M. and Banday, A. J. and Barreiro, R. B. and Eriksen, H. K. and G\'orski, K. M. and Keskitalo, R. and Komatsu, E. and Mart\'{\i}nez-Gonz\'alez, E. and Scott, D. and Vielva, P. and Wehus, I. K.},
  journal = {Phys. Rev. Lett.},
  volume = {128},
  issue = {9},
  pages = {091302},
  numpages = {7},
  year = {2022},
  month = {Mar},
  publisher = {American Physical Society},
  doi = {10.1103/PhysRevLett.128.091302},
  url = {https://link.aps.org/doi/10.1103/PhysRevLett.128.091302}
}

@article{PhysRevD.106.063503,
  title = {Improved constraints on cosmic birefringence from the WMAP and Planck cosmic microwave background polarization data},
  author = {Eskilt, Johannes R. and Komatsu, Eiichiro},
  journal = {Phys. Rev. D},
  volume = {106},
  issue = {6},
  pages = {063503},
  numpages = {10},
  year = {2022},
  month = {Sep},
  publisher = {American Physical Society},
  doi = {10.1103/PhysRevD.106.063503},
  url = {https://link.aps.org/doi/10.1103/PhysRevD.106.063503}
}

@article{ACT,
author = {Louis, Thibaut and Posta, Adrien and Atkins, Zachary and Jense, Hidde and Abril-Cabezas, Irene and Addison, Graeme and Ade, Peter and Aiola, Simone and Alford, Tommy and Alonso, David and Amiri, M. and An, Rui and Austermann, Jason and Barbavara, Eleonora and Battaglia, Nicholas and Battistelli, Elia and Beall, J. and Bean, Rachel and Beheshti, Ali and Zheng, Kaiwen},
year = {2025},
month = {11},
pages = {062},
title = {The Atacama Cosmology Telescope: DR6 power spectra, likelihoods and $\Lambda$CDM parameters},
volume = {2025},
journal = {Journal of Cosmology and Astroparticle Physics},
doi = {10.1088/1475-7516/2025/11/062}
}

@article{pbc3-t52s,
  title = {Cosmic birefringence from the Atacama Cosmology Telescope Data Release 6},
  author = {Diego-Palazuelos, Patricia and Komatsu, Eiichiro},
  journal = {Phys. Rev. D},
  volume = {113},
  issue = {10},
  pages = {L101302},
  numpages = {7},
  year = {2026},
  month = {May},
  publisher = {American Physical Society},
  doi = {10.1103/pbc3-t52s},
  url = {https://link.aps.org/doi/10.1103/pbc3-t52s}
}

@article{Bicepkeck,
doi = {10.3847/1538-4357/acc85c},
url = {https://doi.org/10.3847/1538-4357/acc85c},
year = {2023},
month = {may},
publisher = {The American Astronomical Society},
volume = {949},
number = {2},
pages = {43},
author = {Ade, P. A. R. and Ahmed, Z. and Amiri, M. and Barkats, D. and Thakur, R. Basu and Bischoff, C. A. and Beck, D. and Bock, J. J. and Boenish, H. and Bullock, E. and Buza, V. and IV, J. R. Cheshire and Connors, J. and Cornelison, J. and Crumrine, M. and Cukierman, A. and Denison, E. V. and Dierickx, M. and Duband, L. and Eiben, M. and Fatigoni, S. and Filippini, J. P. and Fliescher, S. and Giannakopoulos, C. and Goeckner-Wald, N. and Goldfinger, D. C. and Grayson, J. and Grimes, P. and Hall, G. and Halal, G. and Halpern, M. and Hand, E. and Harrison, S. and Henderson, S. and Hildebrandt, S. R. and Hubmayr, J. and Hui, H. and Irwin, K. D. and Kang, J. and Karkare, K. S. and Karpel, E. and Kefeli, S. and Kernasovskiy, S. A. and Kovac, J. M. and Kuo, C. L. and Lau, K. and Leitch, E. M. and Lennox, A. and Megerian, K. G. and Minutolo, L. and Moncelsi, L. and Nakato, Y. and Namikawa, T. and Nguyen, H. T. and O’Brient, R. and IV, R. W. Ogburn and Palladino, S. and Petroff, M. and Prouve, T. and Pryke, C. and Racine, B. and Reintsema, C. D. and Richter, S. and Schillaci, A. and Schwarz, R. and Schmitt, B. L. and Sheehy, C. D. and Singari, B. and Soliman, A. and Germaine, T. St. and Steinbach, B. and Sudiwala, R. V. and Teply, G. P. and Thompson, K. L. and Tolan, J. E. and Tucker, C. and Turner, A. D. and Umiltà, C. and Vergès, C. and Vieregg, A. G. and Wandui, A. and Weber, A. C. and Wiebe, D. V. and Willmert, J. and Wong, C. L. and Wu, W. L. K. and Yang, H. and Yoon, K. W. and Young, E. and Yu, C. and Zeng, L. and Zhang, C. and Zhang, S. and BICEP/Keck Collaboration},
title = {BICEP/Keck. XVII. Line-of-sight Distortion Analysis: Estimates of Gravitational Lensing, Anisotropic Cosmic Birefringence, Patchy Reionization, and Systematic Errors},
journal = {The Astrophysical Journal}
}

@misc{kramer2026atacamacosmologytelescopeconstraints,
      title={The Atacama Cosmology Telescope: Constraints on the anisotropic screening and birefringence effects with DR6}, 
      author={Darby M. Kramer and Alexander van Engelen and Frank J. Qu and Christopher Cain and Irene Abril-Cabezas and J. Richard Bond and Vera Gluscevic and Carlos Hervías-Caimapo and Niall MacCrann and Mathew S. Madhavacheril and Jeff McMahon and Toshiya Namikawa and Bruce Partridge and Emmanuel Schaan and Neelima Sehgal and Cristóbal Sifón and Ed Wollack},
      year={2026},
      eprint={2608.27458},
      archivePrefix={arXiv},
      primaryClass={astro-ph.CO},
      url={https://arxiv.org/abs/2608.27458}, 
}

@article{SPTbir,
author = {Balkenhol, Lennart and Coerver, A. and Reichardt, C. and Zebrowski, J.},
year = {2025},
month = {11},
pages = {},
title = {Probing Anisotropic Cosmic Birefringence with Foreground-Marginalised SPT B-mode Likelihoods},
volume = {8},
journal = {The Open Journal of Astrophysics},
doi = {10.33232/001c.147459}
}

@article{Lonappan,
  title = {Constraints on anisotropic cosmic birefringence from CMB $B$-mode polarization},
  author = {Lonappan, A. I. and Keating, B. and Arnold, K.},
  journal = {Phys. Rev. D},
  volume = {112},
  issue = {2},
  pages = {023555},
  numpages = {8},
  year = {2025},
  month = {Jul},
  publisher = {American Physical Society},
  doi = {10.1103/yxmh-rh9z},
  url = {https://link.aps.org/doi/10.1103/yxmh-rh9z}
}

@article{Gruppuso_2012,
doi = {10.1088/1475-7516/2012/02/023},
url = {https://doi.org/10.1088/1475-7516/2012/02/023},
year = {2012},
month = {feb},
publisher = {},
volume = {2012},
number = {02},
pages = {023},
author = {A. Gruppuso and P. Natoli and N. Mandolesi and A. De Rosa and F. Finelli and F. Paci},
title = {WMAP 7 year constraints on CPT violation from large angle CMB anisotropies},
journal = {Journal of Cosmology and Astroparticle Physics}
}


\end{document}